\documentstyle[12pt]{article}
\input{psfig.tex}
\newcommand{\etal}{{\em et\ al.}\ }

\newcommand{\solarm}{$M_{\odot}$\ }

\newcommand{\chisq}{${\chi}^{2}$}

\def\lesssim{\mathrel{\hbox{\rlap{\hbox{\lower4pt\hbox{$\sim$}}}\hbox{$<$}}}}
\def\gtrsim{\mathrel{\hbox{\rlap{\hbox{\lower4pt\hbox{$\sim$}}}\hbox{$>$}}}}

\begin{document}

\vspace{0.1in}

\centerline{                       THE MORPHOLOGY OF TYPE Ia}
\centerline{                       SUPERNOVAE LIGHT CURVES}

\vspace{0.2in}
\centerline{                               Mario Hamuy$^{1,2}$}
\centerline{                              M. M. Phillips$^1$}
\centerline{                           Nicholas B. Suntzeff$^1$}
\centerline{                            Robert A. Schommer$^1$}
\centerline{                                 Jos\'{e} Maza$^{3,4}$}
\centerline{                                 R. C. Smith$^{1,5}$}
\centerline{                                 P. Lira$^{3,6}$}
\centerline{                                 R. Avil\'{e}s$^1$}
\vspace{0.2in}
\noindent $^1$ National Optical Astronomy Observatories$^*$, Cerro Tololo Inter-American Observatory,\\
\noindent Casilla 603, La Serena, Chile\\
\noindent $^2$ University of Arizona, Steward Observatory, Tucson,
Arizona 85721\\
\noindent $^3$ Departamento de Astronom\'{i}a, Universidad de Chile, Casilla 36-D,
Santiago, Chile\\
\noindent $^4$ C\'{a}tedra Presidencial de Ciencias (Chile), 1996-1997.\\
\noindent $^5$ University of Michigan, Department of Astronomy, Michigan 48109-1090\\
\noindent $^6$ Institute for Astronomy, The University of Edinburgh, Royal Observatory, Edinburgh EH9 3HJ, UK\\

\noindent electronic mail: mhamuy@as.arizona.edu, mphillips@noao.edu, nsuntzeff@noao.edu,\\
\noindent rschommer@noao.edu, jmaza@das.uchile.cl, chris@astro.lsa.umich.edu, P.Lira@roe.ac.uk\\
\vspace{0.2in}

\noindent Running Page Head : SNe Ia LIGHT CURVES\\

\noindent Address for proofs:  M. M. Phillips, CTIO, Casilla 603, La Serena, Chile\\

\vspace{0.2in}

\footnoterule

\vspace{0.1in}

\noindent $^*$Cerro Tololo Inter-American Observatory, National
Optical Astronomy Observatories, operated by the Association of Universities
for Research in Astronomy, Inc., (AURA), under cooperative agreement with
the National Science Foundation.\\

\eject
\centerline{                        ABSTRACT}

     We present a family of six BVI template light curves for SNe Ia for days -5 and +80,
based on high-quality data gathered at CTIO.
These templates display a wide range of light
curve morphologies, with initial decline rates of their B light curves
between $\Delta$m$_{15}$(B)=0.87$^{m}$ and 1.93$^{m}$. We use these templates
to study the general morphology of SNe Ia light curves.
We find that several of the main features of the BVI templates  correlate tightly
with $\Delta$m$_{15}$(B).
In particular, the V light curves, which are probably a reasonably good approximation
of the bolometric light curves, display an orderly progression in shapes between
the most-luminous, slowest-declining events and the least-luminous, fastest-declining
SNe.  This supports the idea that the observed spectroscopic and photometric 
sequences of SNe~Ia are due primarily to one parameter.  Nevertheless, SNe
with very similar initial decline rates do show significant differences in their
light curve properties when examined in detail, suggesting the influence of
one or more secondary parameters.

\eject

\section{  Introduction}
     The relatively small dispersion ($\sim$0.6$^{m}$) of the first Hubble diagram
of Type I supernovae (SNe hereafter) constructed by Kowal (1968) revealed the
potential utility of such objects as extragalactic distance indicators. Since then,
considerable effort has 
been devoted to study the photometric properties of these
objects (Leibundgut 1990, van den Bergh \& Pazder 1992, Sandage \& Tammann 1993).
A major difficulty in these studies, however, was caused by the scarcity of
photometric data. An inspection of the atlas of SNe Ia light curves
(Leibundgut \etal 1991) reveals that most of the historical SNe possess fragmentary
data and only a few observations bracket the peaks of the 
light curves. In order
to remedy this situation in 1990 we initiated a photographic survey program, as a
collaboration between the University of Chile and the Cerro Tololo Inter-American
Observatory (CTIO), with the aim to discover young SNe Ia and to populate the
Hubble diagram of these objects in a wide range of redshifts (Hamuy \etal 1993a,
hereafter referred to as Paper I). In the course of 1990-93 this project yielded
high-quality CCD data for $\sim$30 SNe Ia (0.01 $\lesssim$ z $\lesssim$ 0.1) in the
BVI system. Despite our efforts to discover SNe in their rising branch (by monitoring
selected fields regularly twice per month), the rapid evolution of these objects
often prevented their discovery until after maximum light. In fact, out of the $\sim$30
Cal\'{a}n/Tololo SNe~Ia, approximately 
half were discovered near or at maximum light, while
the other half were found within the first 15 days after maximum light (Hamuy \etal
1996b, herefater referred as to paper VII).

     In order to estimate peak magnitudes for those SNe with incomplete data,
the technique of fitting average template light curves has been traditionally
employed (eg. Leibundgut \etal 1991). The most widely used template has been
that calculated by Leibundgut (1988) in the UBVJHK system from the best
observations available at that time. However, recent studies of well-observed
SNe Ia (Phillips \etal 1987; Phillips \etal 1992; Filippenko \etal 1992a; 
Filippenko \etal 1992b; Leibundgut \etal 1993;
Maza \etal 1994, hereafter referred to as Paper II) showed that the light
curves displayed by these objects are not all identical, posing an
additional problem for the determination of the peak magnitudes. The need to
estimate reliable peak magnitudes led us to construct a family of template
curves representing the wide range of light curve morphologies of SNe Ia since,
as demonstrated in Paper I, the error introduced by extrapolating a peak magnitude
using an inappropriate template curve can be substantial. From the best-observed
data available to us we produced six BVI template curves for different
types of SNe Ia, suitable for the analysis of the Cal\'{a}n/Tololo database.
In this paper we present these templates so that they might be
available for other studies.
In Sec. 2 we describe the construction procedure of the templates; in Sec. 3
we describe the properties of the templates; in Sec. 4 we perform
a comparison of the individual properties of the templates and, finally, in Sec. 5
we include a short discussion of the major conclusions that can be drawn from
these templates concerning the physics of SNe~Ia explosions.

\section{  The construction of the templates}

     Since most of the photometric observations of the historical SNe were carried
out with blue photographic plates, the first evidence 
for inhomogeneities within the Ia class was
originally detected in the pg or B light curves. Specifically,
a number of authors (Barbon \etal 1973, Pskovskii 1984, Phillips \etal 1987) noted
that the decline rate of the B light curve during the postmaximum phase varied
substantially from object to object. Recently, Phillips (1993) introduced the parameter
$\Delta$m$_{15}$(B), defined as the amount in magnitudes that the B light curve
decays in the first 15 days after maximum, in order to quantify the decline rate
of the B light curve. The SNe used here to construct the templates (1992bc,
1991T, 1992al, 1992A, 1992bo, 1993H, and 1991bg) were specially selected to span a
range in $\Delta$m$_{15}$(B), between 0.87$^{m}$ and 1.93$^{m}$. With more recent
data obtained at longer wavelengths, it became evident that the inhomogeneities
of SNe Ia are much more pronounced in the I band than in blue light (eg. Suntzeff 1996).
The selected objects are representative of the ample variety of I light curve
morphologies displayed by SNe Ia. Based on these SNe, 
we constructed six BVI template light
curves. Except for SNe 1992bo and 1993H, the light curves of each SN
were sufficiently well-sampled that interpolations could be done with reasonable
confidence. 
Since the photometry for SN 1992bo was somewhat sparser than that for other SNe, we decided
to include 1993H (with a similar $\Delta$m$_{15}$(B)) to complement the light curve of
the former. In general terms, we performed the construction of the templates in
the following manner:\\

\noindent a) We estimated the time of maximum light in B (t$_{0}$$^{B}$) as well
as the BVI magnitudes at t$_{0}$$^{B}$ of each SN by fitting a cubic spline to the
data spanning days -7 to +7 (with respect to t$_{0}$$^{B}$).\\

\noindent b) We defined the time axis such that
t$_{0}$$^{B} \equiv 0$, 
and then shifted each light curve individually along the magnitude axis so that
B=V=I=0.00$^{m}$ at t$_{0}$$^{B}$. Note that this normalization is identical to the
convention adopted by Leibundgut (1988) and {\it preserves the original temporal
separation of the BVI peaks}.\\

\noindent c) We carried out cubic spline fits to the normalized data in the range
-5 $\le$ t$_{0}$$^{B}$ $\le$ +80. We performed the fits separately for various
regions of the light curve, while exercising special care to guarantee the continouity of the
resulting template. We interpolated values with a separation of one day in the range
-5 $\le$ t$_{0}$$^{B}$ $\le$ +32, every 3 days in the range
+35 $\le$ t$_{0}$$^{B}$ $\le$ +50 and every 5 days in the range
+55 $\le$ t$_{0}$$^{B}$ $\le$ +80.

\noindent Next, we summarize the individual details involved in the construction
of each template, presented in order of increasing values of $\Delta$m$_{15}$(B).\\

\subsection{ SN 1992bc}

     We calculated the template for this SN from our own photometry obtained
at CTIO (Paper VII). Since this SN occurred at a redshift
of z=0.02 we started by dividing the timescale of the observed photometry by a 
factor of 1.02 in order
to remove the effect of time dilation. Figure 1
shows the data for SN 1992bc, duly normalized and corrected for time dilation,
along with the template obtained. After calculating the template, we subtracted
the K-terms for SNe Ia calculated by Hamuy \etal (1993b) for the corresponding
redshift. Table 1 (columns 2,3,4) gives the resulting template for zero redshift.
The B template is characterized by $\Delta$m$_{15}$(B)=0.87.

\subsection{ SN 1991T}

      We calculated the template for this SN from a recent reduction of our
own photometry obtained at CTIO (Lira 1995). Due to the proximity of the SN, we did
not apply corrections for time dilation or K terms. Figure 2 shows the normalized data for SN 1991T
along with the template obtained. Table 1 (columns 5,6,7) gives the
resulting template. The B template is characterized by $\Delta$m$_{15}$(B)=0.94.

\subsection{ SN 1992al}

     We calculated the template for this SN from our own photometry obtained at
CTIO (Paper VII). Unfortunately, we could not determine
the peak I magnitude for this SN because the first observation
through this filter started only $\sim$2-3 days after t$_{0}$$^{B}$. To get
around this we adopted I(t$_{0}$$^{B}$=3) = 0.08$^{m}$, based on the
average value yielded by other templates. Inspection of Table 1 shows that
this  is a reasonable assumption since, except for SN 1991bg (the most extreme
of the templates), the normalized I magnitude at day +3 varies in a small range
between 0.06$^{m}$ and 0.11$^{m}$. After 
normalizing the light curves,
we divided the timescale by a factor of 1.015 (since this SN occurred at a 
redshift of z=0.015) in order 
to remove the effect of time dilation. Figure 3
shows the data for SN 1992al, duly normalized and corrected for time dilation, along
with the template obtained. Shown as a dotted line
in the bottom panel of this
figure is an extension to the calculated template I light curve, based on an
average of the remaining templates, which is our best estimate of the behavior
of the I light curve at epochs $-5 \leq$ t$_{0}^{B} \leq +2$.
After calculating the template, we subtracted the K-terms for SNe Ia 
(Hamuy \etal 1993b) 
for the corresponding redshift. Table 1 (columns 8,9,10)
gives the resulting template for zero redshift.
The B template is characterized by $\Delta$m$_{15}$(B)=1.11.

\subsection{ SN 1992A}

      We calculated the template for this SN from our own photometry obtained
at CTIO which will be published elsewhere (Suntzeff \etal 1996). Due to the
proximity of the SN, we did not apply corrections for time dilation or K terms.
Figure 4 shows the normalized data for SN 1992A along with the template
obtained. Table 1 (columns 11,12,13) lists
the resulting template. The B template is characterized by $\Delta$m$_{15}$(B)=1.47.

\subsection{ SNe 1992bo and 1993H}

      Inspection of our photometry for SN 1992bo revealed that this object was
characterized by $\Delta$m$_{15}$(B)$\sim$1.7 (Paper II). Although the first observations
of this SN started several days before maximum light, the sampling of the I light curve
around the secondary maximum did not allow us
to adequately interpolate values for the
construction of the template curve. The only other object with available BVI photometry
and a similar decline rate is SN 1993H.
Fortunately, our followup photometry for this object (Paper VII) started right at
maximum light, and the sampling of the BVI light curves proved to be a good
complement to the observations of SN 1992bo. An additional coincidence between this
pair of SNe was their similar redshifts (z=0.02 for SN 1992bo, and z=0.024 for SN 1993H).
We were therefore able to combine these data without worrying about 
redshift corrections. In doing so,
we used a \chisq minimizing procedure to determine the relative shifts  in the time
and magnitude scales of both objects. Once the two sets of light curves had
been combined, we divided the timescale of both SNe by a factor of 1.02 to remove 
the effect of time dilation. Figure 5 shows the data for SNe 1992bo and 1993H,
duly normalized and corrected for time dilation, along with the template obtained.
After calculating the template, we subtracted the K-terms for SNe Ia calculated
by Hamuy \etal (1993b) for the corresponding redshift. Table 1 (columns 14,15,16)
gives the resulting template for zero redshift. The B light curve is characterized by 
$\Delta$m$_{15}$(B)=1.69.

\subsection{ SN 1991bg}

      We calculated the template for this SN from the photometry published by
Filippenko \etal (1992b) and Leibundgut \etal (1993). These observations
indicate that maximum light was observed in V. However, the
first two observations obtained in B show that the SN was already in the
decline phase. In order to normalize the light curve along the time axis we
choose to adopt the time of the first observation in B as the time of maximum in
this band. With this assumption, the time of maximum in V occurred on  day +2, in
agreement with other SNe Ia (see Table 1). Nevertheless, 
we cannot rule out the possibility that
maximum light in B occurred earlier than t$_{0}$$^{B}$=0. Due to the proximity of the
SN, we did not apply corrections for time dilation or K terms. Figure 6 shows the
normalized data for SN 1991bg along with the template obtained. Table 1
(columns 17,18,19) gives the resulting template. The B template is
characterized by $\Delta$m$_{15}$(B)=1.93.

\section{ General properties of the templates}

     Figure 7 shows all of our BVI templates plotted together to illustrate
the significant differences that occur within this
family of light curves. Despite these  differences,
it is possible to identify a common pattern as well as a number
of key parameters that characterize the shape of the individual templates.
Among these parameters are the time and magnitude at maximum light in each band,
t$_{0}$$^{k}$ and m$_{0}$$^{k}$, where k = B,V,I.
We can also define parameters $\Delta$m$_{T}$(k) which, generalizing Phillips' (1993)
convention, correspond to the amount in magnitudes by which the "k" light curve
decays in the first "T" days after {\it maximum light}.
Figure 8 shows one of the BVI templates with a typical decline rate
(SN 1992al) along with the graphical representation of these parameters.

     The B template light curves are characterized by a rapid rise which
culminates at maximum light. After maximum light the SN displays a fast-decline
phase followed by a slower linear decay in luminosity. 
The two decline phases are
separated by an {\it inflection} point where the curvature of the light curve changes sign.
Depending on the SN, this point is located between
7 and 21 days after maximum. We call this parameter t$_{1}$$^{B}$ which we
define as the time when the first derivative of the B light curve
reaches a maximum.  Another useful time parameter is given by the intersection 
of the straight line that fits the exponential tail and the straight line that
fits the initial decline phase of the B light curve. These two fits are shown as
dotted lines in Figure 8 and we call this point the {\it intersection} parameter, t$_{2}$$^{B}$,
which varies between day +14 and +38 depending on the SN template. This point can
be considered as the onset of the exponential tail.
Another useful difference of these templates is the brightness of the linear tail
(relative to the peak) which we measure here with the parameter $\Delta$m$_{60}$(B).

      The V light curves have a very similar behavior to the B curves, except
that the inflection is much less pronounced  and that the initial postmaximum decline rate is
also smaller. We introduce the parameter $\Delta$m$_{20}$(V) (using a slightly
larger baseline in time than in B to compensate for the slower decline rate) 
in order to quantify the initial decline rate,
and $\Delta$m$_{60}$(V) in order to measure the brightness
of the linear tail. An interesting feature of these templates is that the time of
V maximum occurs 0.5 and 2.5 days later than B maximum, in reasonable agreement
with the finding of Leibundgut (1988) that t$_{0}$$^{V}$ - t$_{0}$$^{B}$ = 2.5 $\pm$ 1.0 days.

     Except in the case of SN 1991bg, the I templates display a significantly more complex
morphology than the B or V light curves. In general terms, the I templates show
a primary maximum which occurs 1-2.5 days {\it before} B maximum; a minimum which occurs
on day 11.5-19 relative to B maximum; a secondary maximum which takes place 18.5-31 days
after B maximum; and a linear decline in magnitude thereafter.
This secondary maximum is observed to be even stronger in the JHK photometry of
SNe~Ia (Elias \etal 1981, 1985).
Since the time of occurence and strength of these features differ significantly within our family
of templates, we use here (t$_{1}$$^{I}$,m$_{1}$$^{I}$) and (t$_{2}$$^{I}$,m$_{2}$$^{I}$)
to define the time and magnitude of the minimum and the secondary maximum of the I template
light curve, respectively. As opposed to the other templates, the 1991bg I template
appears to be characterized by a single maximum. Unfortunately, the observations
of this object started when the SN was already declining in B, so we are unable to
interpret the I maximum as the primary or secondary observed in other SNe Ia.
In fact, as discussed in Sec. 4, it is likely that at such fast decline
rates, the primary and secondary maxima merge into a single maximum.

      Table 2 lists the values of these various parameters for the different
templates. We also include in this table the same parameters for two additional 
well-observed SNe Ia, 1990N and 1994D, which were measured from a new reduction
of the CTIO photometry of SN 1990N (Lira 1995) and a preliminary reduction of
photometry of SN 1994D obtained at CTIO and the Fred Whipple Observatory
(Smith \etal 1996).

\section{ Comparison of the templates}

      Among the various parameters considered here, the initial postmaximum decline
rate of the B light curve is the most extensively studied to date. Phillips (1993)
found that the luminosities of nine nearby, well-observed SNe Ia were tightly correlated with
$\Delta$m$_{15}$(B) in the sense that slow decliners proved to be the most luminous objects.
The $\sim$30 Cal\'{a}n/Tololo SNe Ia provide additional support to the existence of
such a relationship (Hamuy \etal 1996a, hereafter referred to as Paper V). In Figure 9 we show
the six templates on an absolute magnitude scale set by the absolute magnitude-$\Delta$m$_{15}$(B)
relationship given in Paper V.  Nugent \etal (1995) have also shown that certain optical
spectral features correlate with $\Delta$m$_{15}$(B) and were able to explain these as being
due principally to a correlation of the photospheric temperature with $\Delta$m$_{15}$(B).
Given the role that $\Delta$m$_{15}$(B) has played
in the characterization of the intrinsic properties of SNe Ia, we choose to use it here as the
central parameter in our comparison of the templates.

      Figure 10 shows that the time of the {\it inflection} and {\it intersection} points
of the B light curve are both well correlated with $\Delta$m$_{15}$(B),
in the sense that SNe with slower initial decline rates
have a later onset of the exponential tail. Needless to say, both t$_{1}$$^{B}$ and
t$_{2}$$^{B}$ are also very well correlated with each other.  In Figure 11, we show a comparison between the
initial decline rates of the B and V curves, $\Delta$m$_{15}$(B) and $\Delta$m$_{20}$(V), which 
confirms the measurements given in Wells \etal (1994) 
which also indicated that the decline rates in these two colors
are tightly correlated. Next, Figure 12 shows a comparison between $\Delta$m$_{15}$(B)
and the relative brightness of the BVI linear tails. The top panel indicates that there is
no clear relationship between $\Delta$m$_{15}$(B) and $\Delta$m$_{60}$(B). A remarkable
feature in this plot is that, while the 1992bc and 1991T templates have similar
(low) decline rates, these curves differ significantly in the magnitude drop observed
between maximum light and the linear tail (see also Fig. 7). With a somewhat faster
initial decline rate, the SN 1990N template displays an intermediate $\Delta$m$_{60}$(B)
value. Toward higher initial decline rates a regular trend is observed between the templates
made from 1992al, 1994D, 1992A, and 92bo${-}$93H. This trend is broken again by the
1991bg template which displays a relative drop
at day +60 much smaller than the remaining SNe (see also Fig. 7).  In contrast,
Figure 12 shows that the relative brightness of the linear tails in V and I,
$\Delta$m$_{60}$(V) and $\Delta$m$_{60}$(I) are much more tightly correlated with $\Delta$m$_{15}$(B).
This suggests that $\Delta$m$_{60}$(V) and $\Delta$m$_{60}$(I) 
could be well correlated with
each other and, indeed, the bottom panel of
Figure 13 shows an impressive correlation between these two parameters. 
The upper panel of this figure shows only marginal evidence for a correlation
between $\Delta$m$_{60}$(V) and $\Delta$m$_{60}$(B), with SN 1991bg being the most
discrepant point.

We have also compared $\Delta$m$_{15}$(B) with the early-epoch properties of the I light curves.
Figure 14 shows the time and magnitude of the minimum and secondary maximum 
plotted as a function of 
$\Delta$m$_{15}$(B). (Given the ambiguity in the identification of the single maximum
observed for SN 1991bg, we do not include this object in this comparison.)
At first sight, it is difficult to identify clear correlations in these plots. However,
closer inspection reveals that most of the scatter is due to SN 1991T. Leaving aside
this object, the properties of the I templates appear to be generally correlated with
$\Delta$m$_{15}$(B) in the sense that the strength of the minimum and secondary maximum
is greater and occurs later for the slow-declining SNe (and viceversa).

In examining Figures 1-6, it becomes apparent that the second maximum in the I light curve occurs
at nearly the same epoch as the prominent bend in the B light curve, which we have
measured via the {\it intersection} parameter, t$_2^B$.  This is illustrated in Figure 15, where
t$_2^I$ is plotted versus t$_2^B$.  This correlation suggests that the primary and secondary
maxima of the I light curves of the fast-declining SNe~Ia probably merge into a single maximum
or plateau; indeed, if we interpret the initial ``peak'' in the I~light curve of
SN 1991bg as the {\it secondary} maximum, the SN fits quite well on the relation
plotted in Figure 15. This effect would help
to explain the peculiar appearance of the H light curve of the
fast-declining SN 1986G (Frogel \etal 1987) which exhibited a plateau at maximum light,
lasting at least 20 days, rather than the familiar double-peak morphology observed in other
SNe~Ia (Elias \etal 1985).

\section{ Discussion and Conclusions}

The set of templates presented in this paper were selected to be 
representative of the range of observed SNe~Ia light curve morphologies.
During the first $\sim$100 days following maximum light, the bolometric light 
curves of SNe~Ia are fairly well
approximated by the V light curve (e.g., see Suntzeff 1996).
Focussing on the V light curves displayed in Figure~9, there appears to be a
orderly progression from the most-luminous, slowly-declining events like
SNe 1991T and 1992bc, to the least-luminous, fastest-declining SNe such as 92bo/93H and 1991bg.
The strong correlations between $\Delta$m$_{15}$(B) and the two V light
curve decline rate parameters $\Delta$m$_{20}$(V) and $\Delta$m$_{60}$(V)
(see Figures 11 \& 12) strengthens the impression
that the light curve shapes are governed primarily by a single parameter,
consistent with the suggestion by Nugent \etal (1995) that the observed spectroscopic
and photometric sequences of SNe~Ia are due primarily to one parameter. 
These authors concluded that the parameter was most likely 
the mass of radioactive $^{56}$Ni produced in the explosion.
Assuming that the physical conditions in the ejecta of the different SNe during the
final exponential decline phase are similar, the different luminosities during
this phase could be interpreted as reflecting differing $^{56}$Ni masses.
Pinto \& Eastman (1996) have argued that the relative factor is actually the
mass of the progenitor white dwarf star, with masses in the range 0.5-1.4 \solarm
required to explain the range of observed luminosities and decline rates.

Nevertheless, a detailed comparison of the templates of the two slowest-declining, 
most-luminous SNe~Ia, 1992bc and 1991T, shows
that one or more additional parameters probably influence the light curve
shapes.  This is seen most clearly by comparing the B and I light curves of
these two events (see Figure~7).  Although the initial decline rates of 1992bc and 1991T
as measured by both $\Delta$m$_{15}$(B) and $\Delta$m$_{20}$(V) are very similar,
this phase lasts significantly longer in 1992bc (t$^B_2$ occurred 6 days later for 1992bc
than it did for 1991T).  This, in turn leads to the quite different values of 
$\Delta$m$_{60}$(B) observed for these two SNe.  We note that while
the premaximum spectrum of SN 1992bc was very similar to other ``typical'' SNe~Ia 
(Hamuy \etal, in preparation), SN 1991T was an abnormal object with a 
variety of premaximum spectroscopic peculiarities 
(Filippenko \etal 1992a; Phillips \etal 1992). 
Another case in point is SN 1990N, which displayed decline rates
in B and V that were essentially identical to those of SN 1992al, but whose
values of $\Delta$m$_{60}$(B), $\Delta$m$_{60}$(V), and $\Delta$m$_{60}$(I) 
were more similar to those of the slower-declining SN 1991T.  
These small differences could be
due to variations in parameters such as the explosion energy/mechanism and/or the 
distribution of $^{56}$Ni in the ejecta.

Finally, we consider the fastest-declining event in our sample, SN 1991bg.
From Figure~7, one might conclude that this SN does not ``fit in'' well with
the other SNe~Ia since
the final exponential tail of the B light curve sets in very
early, causing the template to cross the other template curves.  This same
property is responsible for the discrepant position of 1991bg
in the top panels of Figures 12 and 13.  However, in Figure~9 which is a
more physically-meaningful plot of the templates, 1991bg does not appear to be
nearly so discordant. Note in particular that the exponential decline phase of the
B light curve does {\it not} cross the other template curves.  
The low luminosity of this SN in the B band and its unusually
red B-V color at maximum (Filippenko \etal 1992b; Leibundgut \etal 1993) were
almost certainly due to a lower than ``normal'' (Nugent \etal 1995) effective temperature.
As mentioned earlier, the bolometric light curves of
all of these SNe are probably closest to the V light curves, and the V light
curve of 1991bg appears to be a fairly smooth extrapolation of the
other template curves (see Figures 7 and 9).  Hence, we conclude that 
the VRI light curve shapes of SN 1991bg are consistent with the
idea that a single major parameter can explain the major characteristics
of the observed sequence of light curve morphologies for SNe Ia.

                              Acknowledgments

This paper has been possible thanks to grant 92/0312 from Fondo Nacional de Ciencias y
Tecnolog\'{i}a (FONDECYT-Chile).
MH acknowledges support provided for this work by the National Science Foundation
through grant number GF-1002-96 from the Association of Universities for Research
in Astronomy, Inc., under NSF Cooperative Agreement No. AST-8947990 and from
Fundaci\'{o}n Andes under project C-12984.
JM and MH acknowledge support by C\'{a}tedra Presidencial de Ciencias 1996-1997.

\eject

\centerline{                                References}

\noindent Barbon, R., Ciatti, F., \& Rosino, L. 1973, A\&AS, 25, 241

\noindent Elias, J.H., Frogel, J.A., Hackwell, J.A., \& Persson, S.E. 1981, ApJ, 251, L13

\noindent Elias, J.H., Matthews, K., Neugebauer, G., \& Persson, S.E. 1985, ApJ, 296, 379

\noindent Filippenko, A.V., \etal 1992a, ApJ, 384, L15

\noindent Filippenko, A.V., \etal 1992b, AJ, 104, 1543

\noindent Frogel, J.A., Gregory, B., Kawara, K., Laney, D., Phillips, M.M.,
     Terndrup, D., Vrba, F., \& Whitford, A.E. 1987, ApJ, 315, L129

\noindent Hamuy, M., \etal 1993a, AJ, 106, 2392 (Paper I)

\noindent Hamuy, M., Phillips, M.M., Wells, L.A., \& Maza, J. 1993b, PASP, 105, 787

\noindent Hamuy, M., Phillips, M.M., Schommer, R.A., Suntzeff, N.B., Maza, J.,
     \& Avil\'{e}s, R. 1996a, AJ, this volume (Paper V)

\noindent Hamuy, M., \etal 1996b, AJ, this volume (Paper VII)

\noindent Kowal, C.T. 1968, AJ, 73, 1021

\noindent Leibundgut, B. 1988, Ph.D. thesis, University of Basel

\noindent Leibundgut, B. 1990, in {\it Supernovae}, edited by S.E. Woosley (Springer, Berlin), p. 751

\noindent Leibundgut, B., Tammann, G.A., Cadonau, R., \& Cerrito, D. 1991, A\&AS, 89, 537

\noindent Leibundgut, B., \etal 1993, AJ, 105, 301

\noindent Lira, P. 1995, MS Thesis, Universidad de Chile

\noindent Maza, J., Hamuy, M., Phillips, M.M., Suntzeff, N.B., \& Avil\'{e}s, R. 1994,
     ApJ, 424, L107 (Paper II)

\noindent Nugent, P., Phillips, M., Baron, E., Branch, D., \& Hauschildt, P. 1995,
     ApJ, 455, L147

\noindent Phillips, M.M., \etal 1987, PASP, 99, 592 

\noindent Phillips, M.M., Wells, L.A., Suntzeff, N.B., Hamuy, M., Leibundgut, B., 
     Kirshner, R.P., \& Foltz, C.B. 1992, AJ, 103, 1632

\noindent Phillips, M.M. 1993, ApJ, 413, L105

\noindent Pinto, P.A., \& Eastman, R.G. 1996, in preparation

\noindent Pskovskii, Y.P. 1984, SvA, 28, 658

\noindent Sandage, A., \& Tammann, G.A. 1993, ApJ, 415, 1

\noindent Smith, R.C., \etal 1996, in preparation

\noindent Suntzeff, N.B. 1996, in {\it Supernovae and Supernova Remnants}, IAU Colloquium 145,
     ed. R. McCray (Cambridge University Press, Cambridge), in press

\noindent Suntzeff, N.B., \etal 1996, in preparation

\noindent van den Bergh, S., \& Pazder, J. 1992, ApJ, 390, 34

\noindent Wells, L.A., \etal 1994, AJ, 108, 2233
\eject

\centerline{                             Figure Captions}

\noindent Figure 1. The B (top), V (middle), and I (bottom) light curves of SN 1992bc,
duly normalized and corrected for time dilation, along with the calculated
template. The B template is characterized by $\Delta$m$_{15}$(B)=0.87$^{m}$.

\noindent Figure 2. The B (top), V (middle), and I (bottom) light curves of SN 1991T,
duly normalized, along with the calculated
template. The B template is characterized by $\Delta$m$_{15}$(B)=0.94$^{m}$.

\noindent Figure 3. The B (top), V (middle), and I (bottom) light curves of SN 1992al,
duly normalized and corrected for time dilation, along with the calculated
template. Shown as dotted lines in the bottom panel is an extension to the calculated
template I light curve, based on an average of the remaining templates. 
The B template is characterized by $\Delta$m$_{15}$(B)=1.11$^{m}$.

\noindent Figure 4. The B (top), V (middle), and I (bottom) light curves of SN 1992A,
duly normalized, along with the calculated
template. The B template is characterized by $\Delta$m$_{15}$(B)=1.47$^{m}$.

\noindent Figure 5. The B (top), V (middle), and I (bottom) light curves of SNe 1992bo
(open circles) and 1993H (crosses), duly normalized and corrected for time dilation,
along with the calculated template. The B template is characterized by $\Delta$m$_{15}$(B)=1.69$^{m}$.

\noindent Figure 6. The B (top), V (middle), and I (bottom) light curves of SN 1991bg,
duly normalized, along with the calculated template. The B template is characterized
by $\Delta$m$_{15}$(B)=1.93$^{m}$.

\noindent Figure 7. The comparison of the six template B (top), V (middle), and I (bottom)
light curves of SNe Ia, all normalized such that m$^{k}$(t$_{0}$$^{B}$)=0 in all three
bands (k=B,V,I).

\noindent Figure 8. The template B (top), V (middle), and I (bottom) light curves of SN 1992al.
Also shown are the graphical representations of the key parameters defined here in order to
characterize the shape of the individual templates.

\noindent Figure 9. The six templates B (top), V (middle), and I (bottom) light curves of SNe Ia,
on the absolute magnitude scale set by the peak luminosity-$\Delta$m$_{15}$(B) relationship
of Paper V. The peak absolute magnitudes for the five templates with 
0.87 $\leq$ $\Delta$m$_{15}$(B) $\leq$ 1.69 were calculated using the ``low-extinction''
fits given in Table~3 of Paper~V.  The SN 1991bg template is plotted at the peak absolute
magnitudes given in Table~1 of Paper~V for the fast-declining event SN 1992K.

\noindent Figure 10. (top) The {\it inflection} point of the B light curve, t$_{1}$$^{B}$,
as a function of $\Delta$m$_{15}$(B). (bottom) The {\it intersection} point of the
B light curve, t$_{2}$$^{B}$, as a function of $\Delta$m$_{15}$(B).

\noindent Figure 11. The initial decline rate of the V template light curves, $\Delta$m$_{20}$(V),
plotted as a function of the initial decline rate of the B template light curves, $\Delta$m$_{15}$(B).

\noindent Figure 12. Comparison of the initial decline rate of the B light curve, $\Delta$m$_{15}$(B),
and the brightness of the linear tail (relative to maximum light) of the B (top), V (middle),
and I (bottom) template light curves.

\noindent Figure 13. (top) The brightness of the linear tail (relative to maximum light) of the
B template light curves, plotted as a function of the brightness of the linear tail
(relative to maximum light) of the V template light curves. (bottom) The same plot for the
I template curves.

\noindent Figure 14. (top) The time and magnitude of the minimum 
of the I template light curves, plotted as a function of $\Delta$m$_{15}$(B).
(bottom) The time and magnitude of the secondary maximum
of the I template light curves, plotted as a function of $\Delta$m$_{15}$(B).

\noindent Figure 15. The time of the secondary maximum of the 
I template light curves, t$^I_2$, plotted as a function of the time
of occurrence of the bend in the B light curves as measured by the
{\it intersection} parameter, t$^B_2$.

\pagestyle{empty}
\eject
\evensidemargin=-0.80in
\oddsidemargin=-0.80in

\tiny
\begin{tabular}{lcccccccccccccccccl}
\multicolumn{19}{c}{\bf Table 1. Templates for Type Ia Supernovae} \\
&&&&&&&&&&&&&&&&&&\\
(1) & (2) & (3) & (4) & (5) & (6) & (7) & (8) & (9) & (10) & (11) & (12) & (13) & (14) & (15) & (16) & (17) & (18) & (19) \\
&&&&&&&&&&&&&&&&&&\\
\hline\hline\\
&&1992bc&&&1991T&&&1992al&&&1992A&&&92bo${-}$93H&&&1991bg&\\
t-t$_{0}$$^{B}$ & B & V & I & B & V & I & B & V & I & B & V & I & B & V & I & B & V & I \\
days&mag&mag&mag&mag&mag&mag&mag&mag&mag&mag&mag&mag&mag&mag&mag&mag&mag&mag\\
\hline\\
-5.0&0.25&0.27& 0.07&0.14& 0.22& 0.09&0.20&0.23& 0.11&0.30& 0.31& 0.06&0.34&0.34& 0.10&....& ....& ....\\
-4.0&0.16&0.18& 0.03&0.09& 0.15& 0.05&0.13&0.15& 0.05&0.19& 0.21&-0.02&0.21&0.23& 0.03&....& ....& ....\\
-3.0&0.10&0.11&-0.01&0.05& 0.10& 0.02&0.07&0.09& 0.00&0.10& 0.14&-0.05&0.12&0.14&-0.01&....& ....& ....\\
-2.0&0.06&0.06&-0.02&0.02& 0.06&-0.00&0.03&0.04&-0.01&0.04& 0.08&-0.05&0.06&0.08&-0.02&....& ....& ....\\
-1.0&0.02&0.03&-0.02&0.00& 0.02&-0.00&0.00&0.01&-0.01&0.01& 0.03&-0.03&0.01&0.04&-0.02&....& ....& ....\\
 0.0&0.00&0.00& 0.00&0.00& 0.00& 0.00&0.00&0.00& 0.00&0.00& 0.00& 0.00&0.00&0.00& 0.00&0.00& 0.00& 0.00\\
 1.0&0.01&0.00& 0.02&0.00&-0.01& 0.01&0.01&0.00& 0.02&0.03&-0.02& 0.02&0.01&0.00& 0.04&0.05&-0.04&-0.04\\
 2.0&0.02&0.01& 0.06&0.02&-0.02& 0.03&0.03&0.01& 0.05&0.07&-0.03& 0.04&0.04&0.01& 0.05&0.10&-0.05&-0.07\\
 3.0&0.04&0.03& 0.11&0.05&-0.02& 0.06&0.05&0.02& 0.08&0.14&-0.02& 0.06&0.10&0.03& 0.09&0.19&-0.04&-0.09\\
 4.0&0.08&0.05& 0.15&0.08&-0.01& 0.10&0.10&0.04& 0.12&0.21& 0.01& 0.10&0.17&0.07& 0.14&0.31& 0.00&-0.10\\
 5.0&0.13&0.08& 0.20&0.13& 0.01& 0.14&0.15&0.06& 0.16&0.29& 0.04& 0.17&0.26&0.10& 0.20&0.46& 0.07&-0.09\\
 6.0&0.18&0.12& 0.26&0.18& 0.03& 0.19&0.22&0.09& 0.21&0.38& 0.09& 0.25&0.37&0.15& 0.25&0.65& 0.15&-0.07\\
 7.0&0.24&0.16& 0.33&0.25& 0.07& 0.26&0.29&0.13& 0.25&0.47& 0.15& 0.33&0.48&0.21& 0.31&0.85& 0.26&-0.04\\
 8.0&0.31&0.22& 0.39&0.32& 0.10& 0.31&0.38&0.18& 0.32&0.57& 0.22& 0.42&0.62&0.27& 0.36&1.06& 0.37&-0.01\\
 9.0&0.37&0.27& 0.46&0.39& 0.14& 0.35&0.46&0.23& 0.43&0.68& 0.30& 0.49&0.75&0.34& 0.41&1.25& 0.50& 0.04\\
10.0&0.45&0.31& 0.53&0.47& 0.18& 0.38&0.57&0.28& 0.51&0.79& 0.38& 0.53&0.91&0.40& 0.45&1.42& 0.64& 0.09\\
11.0&0.53&0.37& 0.58&0.56& 0.23& 0.39&0.67&0.34& 0.57&0.92& 0.45& 0.54&1.06&0.47& 0.47&1.57& 0.76& 0.16\\
12.0&0.61&0.42& 0.63&0.64& 0.28& 0.40&0.78&0.38& 0.61&1.05& 0.52& 0.54&1.24&0.55& 0.49&1.68& 0.89& 0.22\\
13.0&0.70&0.48& 0.68&0.74& 0.34& 0.41&0.89&0.44& 0.64&1.19& 0.59& 0.52&1.40&0.62& 0.49&1.79& 1.01& 0.29\\
14.0&0.78&0.53& 0.73&0.84& 0.39& 0.40&1.01&0.51& 0.65&1.33& 0.65& 0.50&1.55&0.69& 0.49&1.87& 1.12& 0.36\\
15.0&0.87&0.58& 0.75&0.94& 0.45& 0.39&1.11&0.58& 0.66&1.47& 0.71& 0.46&1.69&0.77& 0.47&1.93& 1.21& 0.44\\
16.0&0.96&0.64& 0.78&1.05& 0.51& 0.38&1.23&0.64& 0.65&1.62& 0.77& 0.42&1.84&0.87& 0.45&1.98& 1.30& 0.52\\
17.0&1.05&0.69& 0.79&1.15& 0.56& 0.37&1.34&0.70& 0.64&1.75& 0.82& 0.38&1.96&0.96& 0.44&2.02& 1.37& 0.60\\
18.0&1.14&0.74& 0.81&1.26& 0.62& 0.37&1.47&0.76& 0.62&1.88& 0.88& 0.35&2.09&1.04& 0.43&2.05& 1.44& 0.69\\
19.0&1.24&0.78& 0.82&1.36& 0.67& 0.36&1.58&0.83& 0.60&2.01& 0.94& 0.33&2.22&1.11& 0.43&2.09& 1.50& 0.78\\
20.0&1.35&0.82& 0.81&1.47& 0.72& 0.35&1.69&0.89& 0.58&2.12& 1.01& 0.32&2.32&1.21& 0.44&2.14& 1.55& 0.85\\
21.0&1.45&0.87& 0.81&1.58& 0.77& 0.34&1.81&0.94& 0.56&2.23& 1.09& 0.32&2.42&1.31& 0.46&2.18& 1.59& 0.92\\
22.0&1.55&0.91& 0.79&1.69& 0.83& 0.34&1.92&1.00& 0.54&2.34& 1.20& 0.34&2.52&1.37& 0.50&2.22& 1.63& 0.98\\
23.0&1.65&0.95& 0.78&1.80& 0.88& 0.34&2.02&1.07& 0.53&2.44& 1.29& 0.37&2.61&1.45& 0.56&2.27& 1.67& 1.04\\
24.0&1.74&1.00& 0.76&1.90& 0.93& 0.34&2.12&1.13& 0.51&2.54& 1.37& 0.42&2.69&1.51& 0.62&2.30& 1.72& 1.11\\
25.0&1.83&1.06& 0.75&1.99& 0.99& 0.35&2.22&1.19& 0.50&2.63& 1.46& 0.49&2.75&1.58& 0.70&2.33& 1.77& 1.17\\
26.0&1.92&1.10& 0.73&2.07& 1.04& 0.35&2.32&1.25& 0.50&2.73& 1.54& 0.57&2.82&1.64& 0.80&2.36& 1.82& 1.23\\
27.0&2.02&1.16& 0.72&2.14& 1.09& 0.36&2.40&1.30& 0.50&2.81& 1.61& 0.65&2.88&1.70& 0.89&2.38& 1.86& 1.29\\
28.0&2.10&1.21& 0.71&2.21& 1.14& 0.38&2.48&1.36& 0.51&2.89& 1.68& 0.74&2.92&1.76& 0.94&2.41& 1.92& 1.35\\
29.0&2.19&1.27& 0.70&2.27& 1.19& 0.40&2.55&1.41& 0.53&2.94& 1.74& 0.83&2.96&1.82& 1.03&2.44& 1.97& 1.41\\
30.0&2.27&1.30& 0.69&2.33& 1.24& 0.42&2.63&1.46& 0.55&3.00& 1.80& 0.93&2.98&1.86& 1.10&2.46& 2.02& 1.47\\
31.0&2.34&1.36& 0.69&2.38& 1.29& 0.45&2.69&1.50& 0.58&3.03& 1.86& 1.01&3.03&1.92& 1.18&2.49& 2.07& 1.53\\
32.0&2.42&1.41& 0.69&2.43& 1.33& 0.48&2.74&1.55& 0.62&3.06& 1.91& 1.09&3.07&1.97& 1.25&2.52& 2.12& 1.59\\
35.0&2.62&1.56& 0.74&2.56& 1.47& 0.60&2.90&1.70& 0.77&3.14& 2.05& 1.27&3.16&2.12& 1.45&2.60& 2.26& 1.76\\
38.0&2.77&1.70& 0.82&2.66& 1.59& 0.75&3.02&1.83& 0.93&3.22& 2.16& 1.43&3.24&2.25& 1.63&2.68& 2.40& 1.93\\
41.0&2.91&1.84& 0.98&2.76& 1.71& 0.90&3.11&1.95& 1.11&3.29& 2.26& 1.58&3.32&2.38& 1.82&2.76& 2.54& 2.09\\
44.0&3.03&1.98& 1.15&2.84& 1.83& 1.08&3.19&2.07& 1.28&3.36& 2.35& 1.73&3.39&2.49& 1.97&2.84& 2.66& 2.25\\
47.0&3.12&2.09& 1.36&2.90& 1.93& 1.26&3.27&2.18& 1.46&3.41& 2.43& 1.88&3.45&2.59& 2.14&2.90& 2.77& 2.40\\
50.0&3.20&2.21& 1.56&2.96& 2.04& 1.43&3.34&2.28& 1.64&3.45& 2.51& 2.03&3.50&2.68& 2.29&2.98& 2.88& 2.55\\
55.0&3.29&2.39& 1.87&3.05& 2.21& 1.69&3.39&2.45& 1.87&3.51& 2.65& 2.27&3.58&2.83& 2.54&3.09& 3.04& 2.78\\
60.0&3.36&2.54& 2.09&3.12& 2.37& 1.92&3.47&2.59& 2.09&3.56& 2.78& 2.50&3.64&2.96& 2.76&3.19& 3.19& 3.00\\
65.0&3.42&2.69& 2.32&3.19& 2.51& 2.11&3.53&2.74& 2.30&3.61& 2.92& 2.73&3.71&3.10& 3.00&3.30& 3.33& 3.22\\
70.0&3.48&2.83& 2.53&3.29& 2.64& 2.32&3.60&2.87& 2.51&3.66& 3.06& 2.94&3.77&3.24& 3.21&3.40& 3.47& 3.41\\
75.0&3.55&2.96& 2.74&3.38& 2.75& 2.54&3.66&2.98& 2.72&3.72& 3.20& 3.15&3.86&3.39& 3.44&3.49& 3.62& 3.59\\
80.0&3.62&3.08& 2.94&3.46& 2.86& 2.73&3.73&3.11& 2.92&3.80& 3.34& 3.35&3.96&3.55& 3.68&3.58& 3.76& 3.76\\
\hline\hline\\
             
\end{tabular}
\eject

\tiny
\evensidemargin=-0.7in
\oddsidemargin=-0.7in
\begin{tabular}{ccccccccccccccll}
\multicolumn{16}{c}{\bf Table 2. Parameters for Template Light Curves     }\\
&&&&&&&&&&&&&&&\\
&&&&&&&&&&&&&&&\\
\hline\hline\\
 SN & $\Delta$m$_{15}$(B) & t$_{1}$$^{B}$ & t$_{2}$$^{B}$ & $\Delta$m$_{60}$(B) & t$_{0}$$^{V}$  & m$_{0}$$^{V}$ & $\Delta$m$_{20}$(V) & $\Delta$m$_{60}$(V) & t$_{0}$$^{I}$ & m$_{0}$$^{I}$ & t$_{1}$$^{I}$ & m$_{1}$$^{I}$ & t$_{2}$$^{I}$ & m$_{2}$$^{I}$ & $\Delta$m$_{60}$(I) \\
&mag&days&days&mag&days&mag&mag&mag&days&mag&days&mag&days&mag&mag\\
\hline
&&&&&&&&&&&&&&&\\
92bc         & 0.87 & 20.6 & 38.0 & 3.36 & 0.5 &  0.00 & 0.85 & 2.56 & -1.5 & -0.02  & 19.0 & 0.82 & 31.0 &  0.69 & 2.04\\
91T          & 0.94 & 19.0 & 32.0 & 3.12 & 2.5 & -0.02 & 0.88 & 2.46 & -1.0 &  0.00  & 13.0 & 0.41 & 22.5 &  0.34 & 1.88\\
92al         & 1.11 & 16.9 & 33.0 & 3.47 & 0.5 &  0.00 & 0.92 & 2.61 & -1.5 & -0.01  & 15.0 & 0.66 & 26.0 &  0.50 & 2.04\\
92A          & 1.47 & 14.1 & 28.5 & 3.56 & 2.0 & -0.03 & 1.23 & 2.87 & -2.5 & -0.05  & 11.5 & 0.54 & 20.5 &  0.32 & 2.44\\
92bo${-}$93H & 1.69 & 11.6 & 25.0 & 3.64 & 0.5 &  0.00 & 1.26 & 2.97 & -1.5 & -0.02  & 13.0 & 0.49 & 18.5 &  0.43 & 2.71\\
91bg         & 1.93 &  7.0 & 14.0 & 3.19 & 2.0 & -0.05 & 1.68 & 3.30 & 4.0? & -0.10? & ...  & ...  & 4.0? & -0.10 & 3.28\\
&&&&&&&&&&&&&&&\\
90N          & 1.07 & 15.8 & 34.0 & 3.24 & 2.5 & -0.01 & 0.90 & 2.42 & -2.1 & -0.02  & 16.5 & 0.58 & ...  &  ...  & 1.72\\
94D          & 1.32 & 15.6 & 28.5 & 3.54 & 1.0 &  0.00 & 1.14 & 2.81 & -2.7 & -0.04  & 11.9 & 0.68 & 20.7 &  0.44 & 2.51\\
&&&&&&&&&&&&&&&\\
\hline\hline\\
\end{tabular}

\begin{figure}
\psfull
%\psdraft
\centerline{\psfig{figure=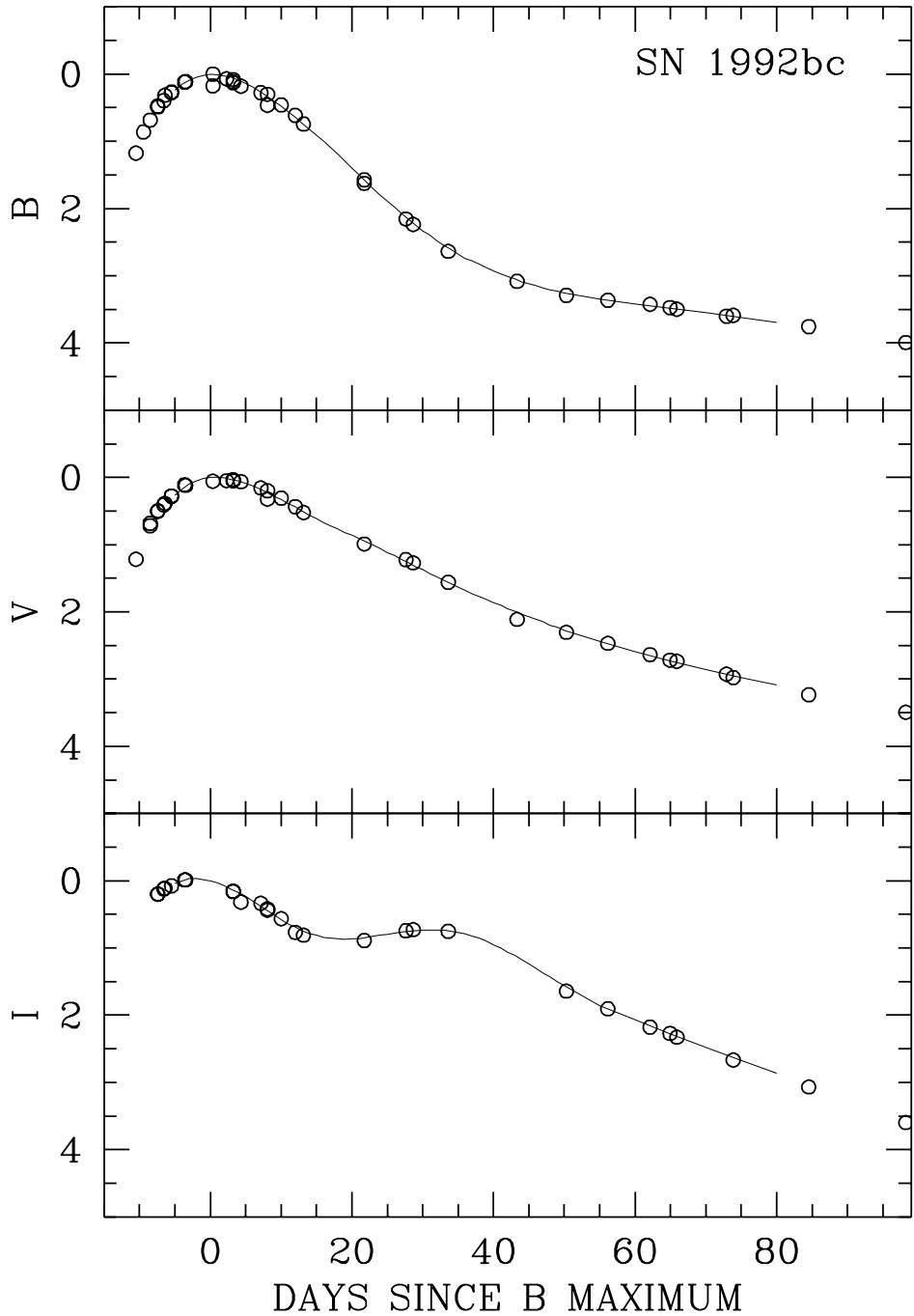}}
\caption{The B (top), V (middle), and I (bottom) light curves of SN 1992bc,
duly normalized and corrected for time dilation, along with the calculated
template. The B template is characterized by $\Delta$m$_{15}$(B)=0.87$^{m}$.}
\end{figure}

\begin{figure}
\psfull
%\psdraft
\centerline{\psfig{figure=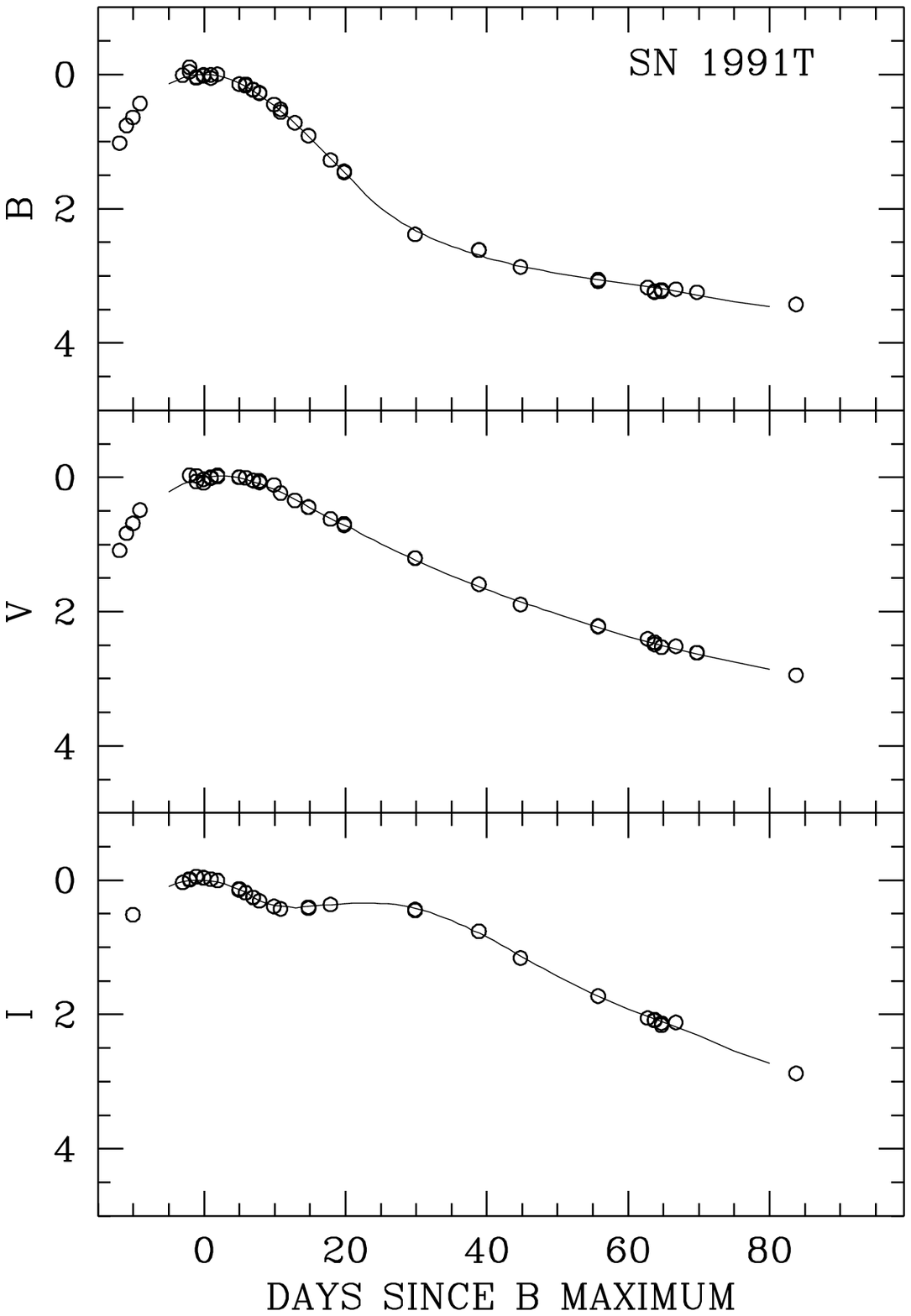}}
\caption{The B (top), V (middle), and I (bottom) light curves of SN 1991T,
duly normalized, along with the calculated
template. The B template is characterized by $\Delta$m$_{15}$(B)=0.94$^{m}$.}
\end{figure}

\begin{figure}
\psfull
%\psdraft
\centerline{\psfig{figure=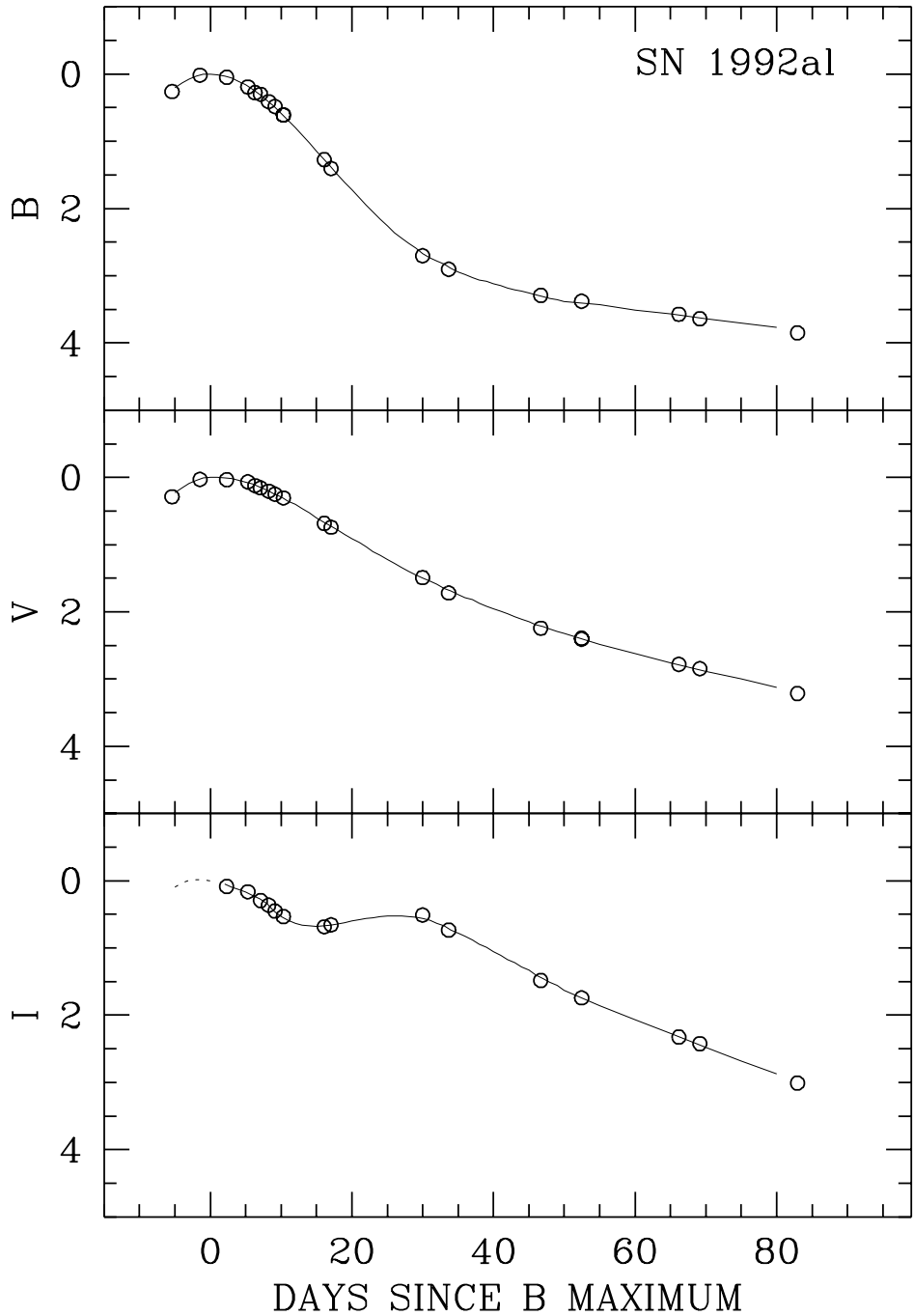}}
\caption{The B (top), V (middle), and I (bottom) light curves of SN 1992al,
duly normalized and corrected for time dilation, along with the calculated
template. Shown as dotted lines in the bottom panel is an extension to the calculated
template I light curve, based on an average of the remaining templates.
The B template is characterized by $\Delta$m$_{15}$(B)=1.11$^{m}$.}
\end{figure}

\begin{figure}
\psfull
%\psdraft
\centerline{\psfig{figure=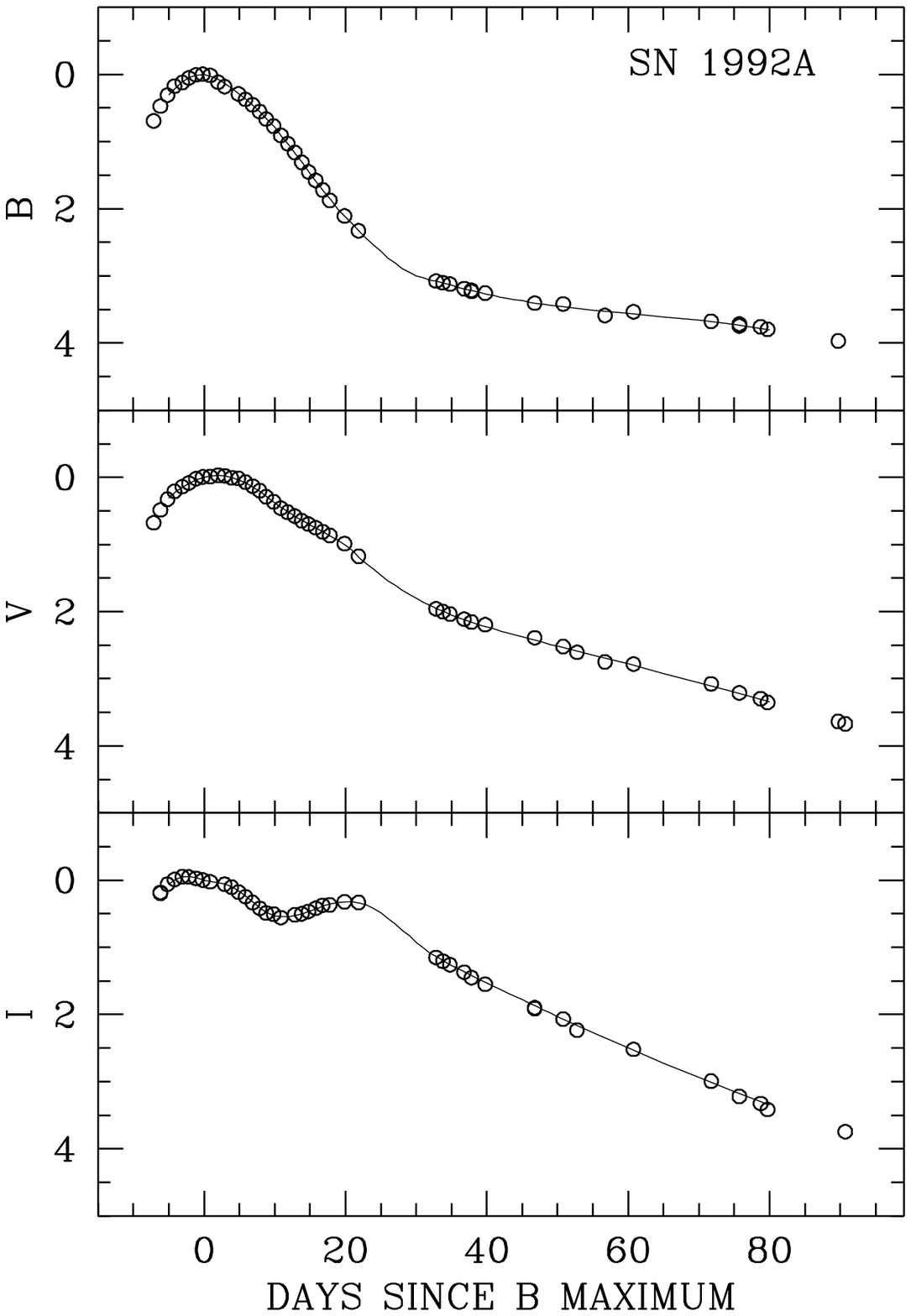}}
\caption{The B (top), V (middle), and I (bottom) light curves of SN 1992A,
duly normalized, along with the calculated
template. The B template is characterized by $\Delta$m$_{15}$(B)=1.47$^{m}$.}
\end{figure}

\begin{figure}
\psfull
%\psdraft
\centerline{\psfig{figure=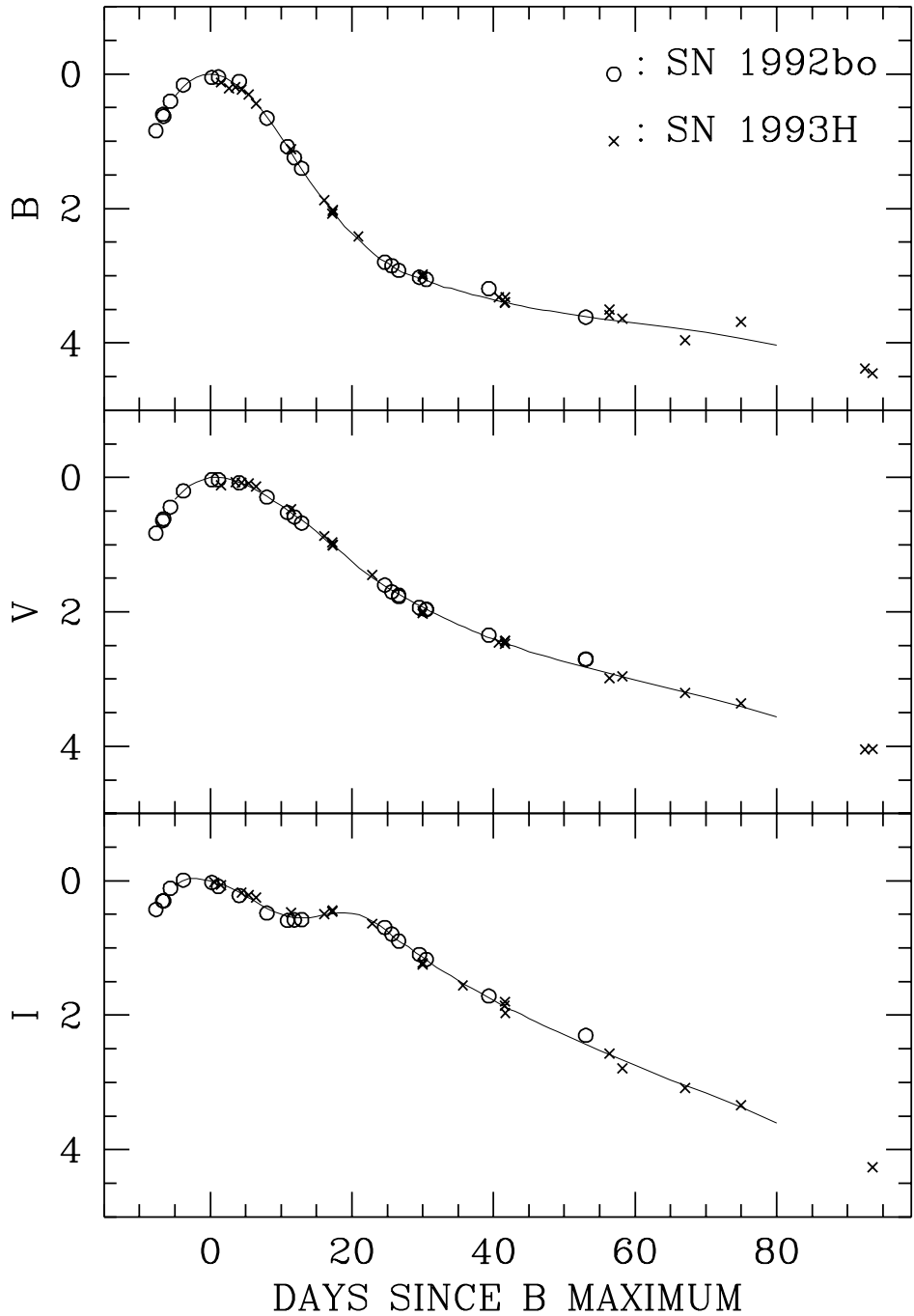}}
\caption{The B (top), V (middle), and I (bottom) light curves of SNe 1992bo
(open circles) and 1993H (crosses), duly normalized and corrected for time dilation,
along with the calculated template. The B template is characterized by $\Delta$m$_{15}$(B)=1.69$^{m}$.}
\end{figure}

\begin{figure}
\psfull
%\psdraft
\centerline{\psfig{figure=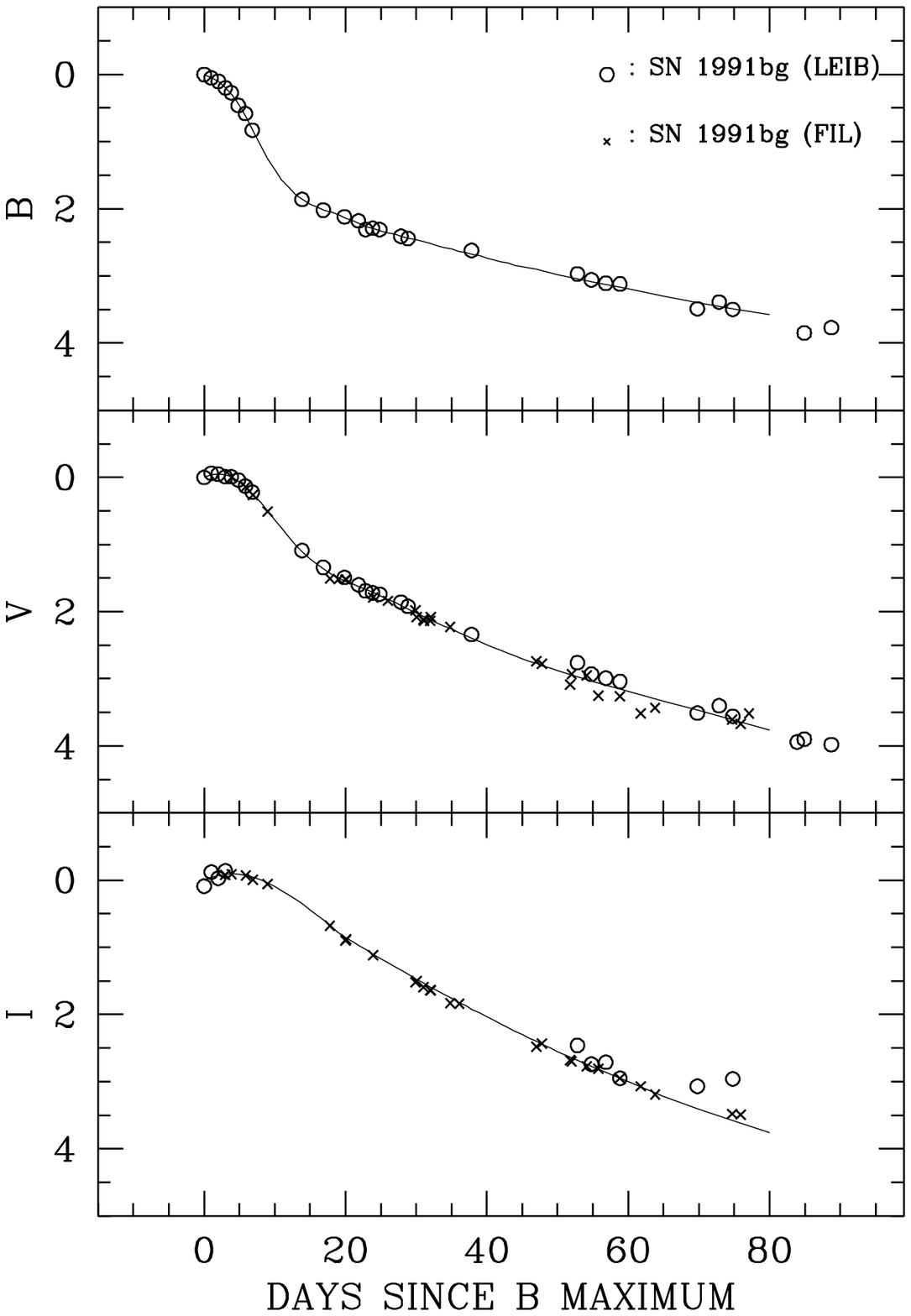}}
\caption{The B (top), V (middle), and I (bottom) light curves of SN 1991bg,
duly normalized, along with the calculated template. The B template is characterized
by $\Delta$m$_{15}$(B)=1.93$^{m}$.}
\end{figure}

\begin{figure}
\psfull
%\psdraft
\centerline{\psfig{figure=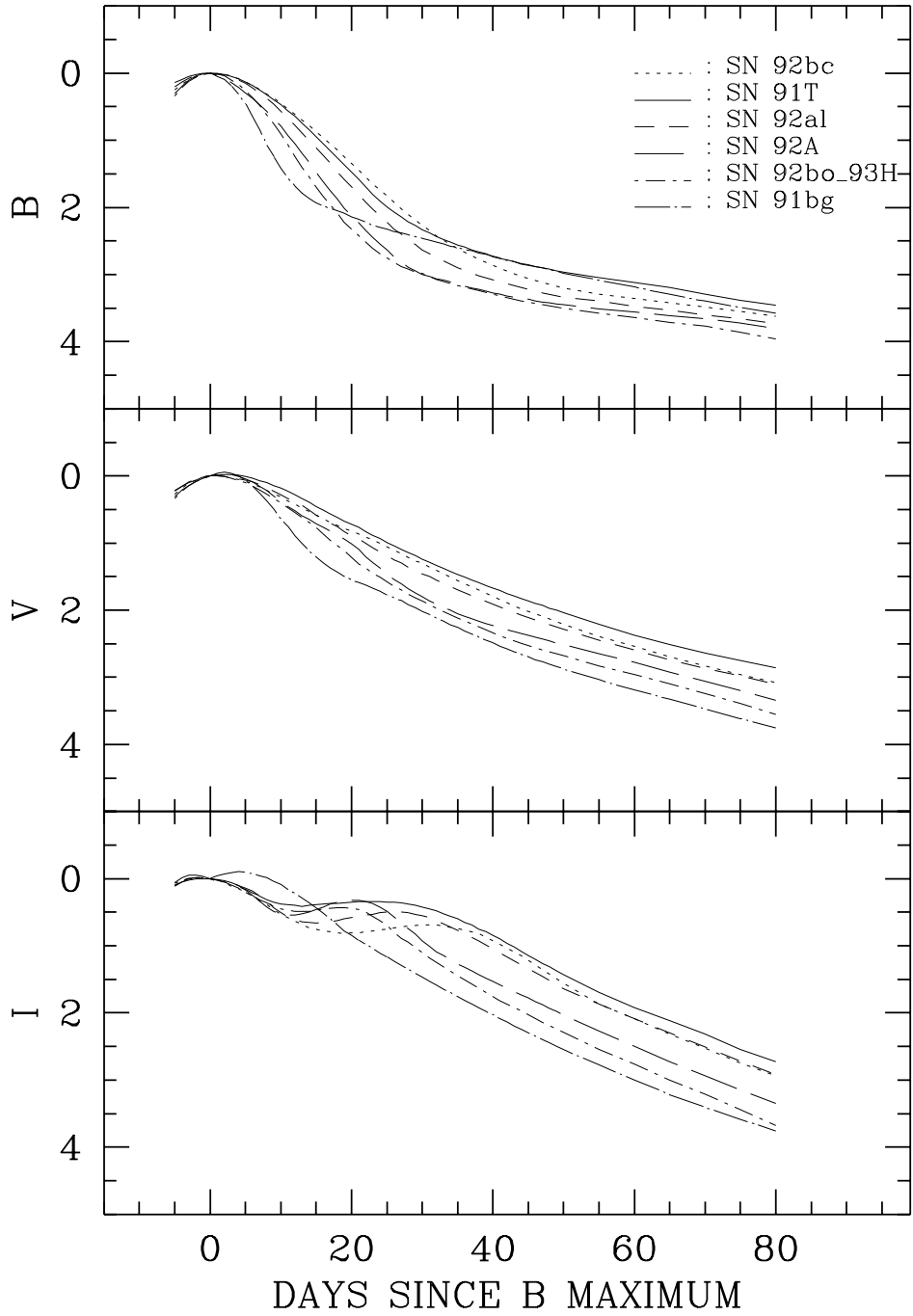}}
\caption{ The comparison of the six template B (top), V (middle), and I (bottom)
light curves of SNe Ia, all normalized such that m$^{k}$(t$_{0}$$^{B}$)=0 in all three
bands (k=B,V,I).}
\end{figure}

\begin{figure}
\psfull
%\psdraft
\centerline{\psfig{figure=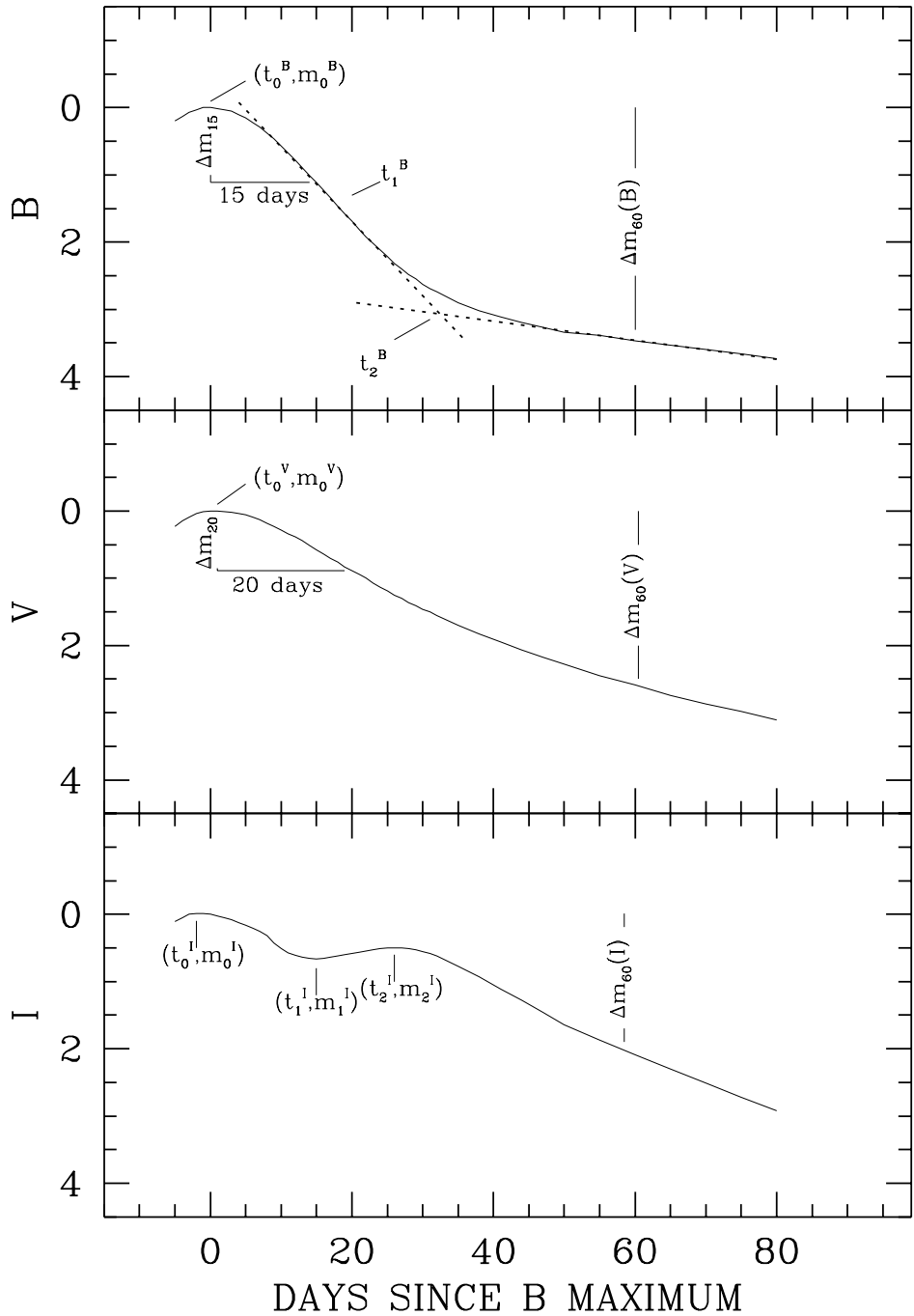}}
\caption{The template B (top), V (middle), and I (bottom) light curves of SN 1992al.
Also shown are the graphical representations of the key parameters defined here in order to
characterize the shape of the individual templates.}
\end{figure}

\begin{figure}
\psfull
%\psdraft
\centerline{\psfig{figure=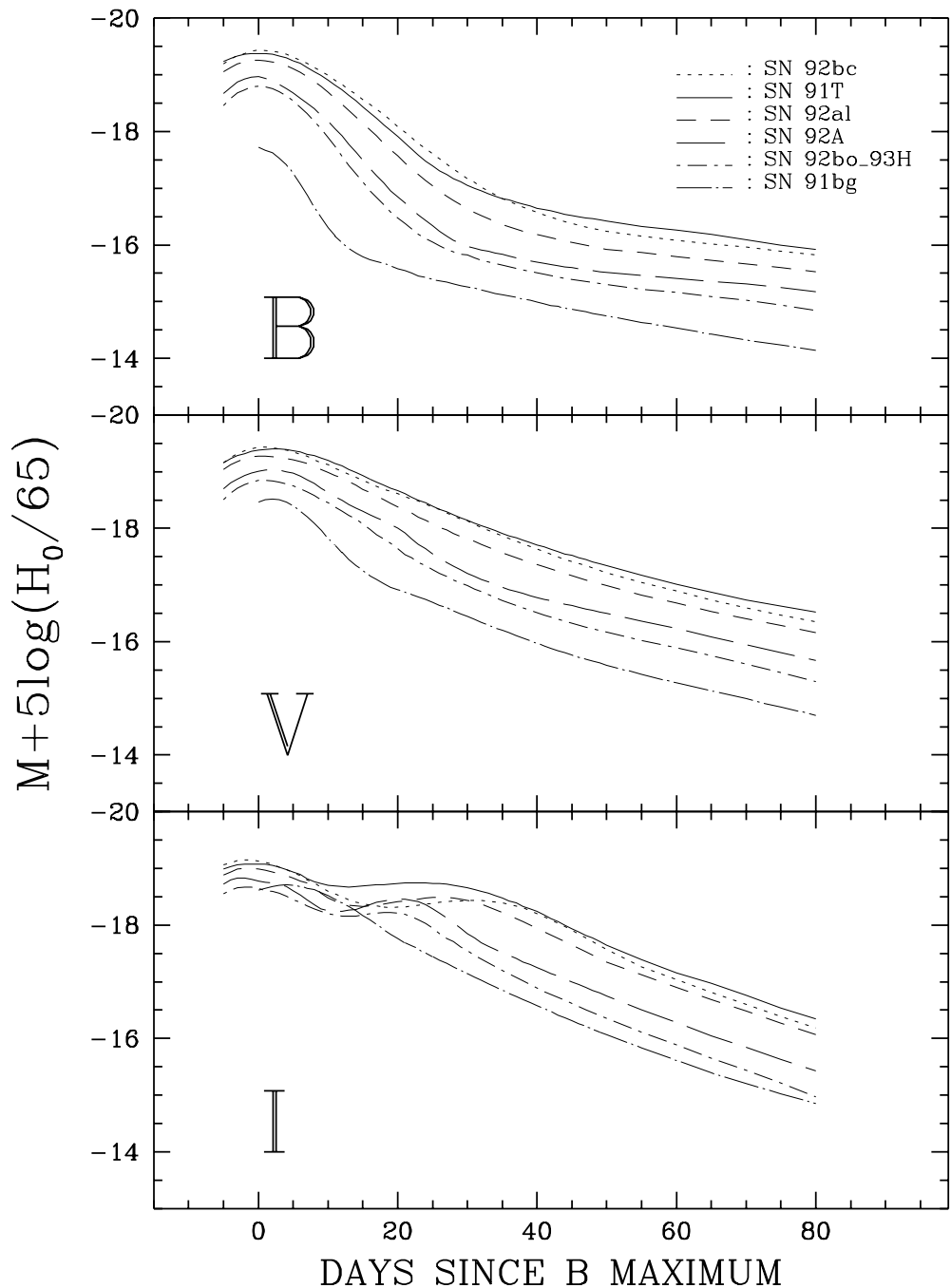}}
\caption{The six templates B (top), V (middle), and I (bottom) light curves of SNe Ia,
on the absolute magnitude scale set by the peak luminosity-$\Delta$m$_{15}$(B) relationship
of Paper V. The peak absolute magnitudes for the five templates with
0.87 $\leq$ $\Delta$m$_{15}$(B) $\leq$ 1.69 were calculated using the ``low-extinction''
fits given in Table~3 of Paper~V.  The SN 1991bg template is plotted at the peak absolute
magnitudes given in Table~1 of Paper~V for the fast-declining event SN 1992K.}
\end{figure}

\begin{figure}
\psfull
%\psdraft
\centerline{\psfig{figure=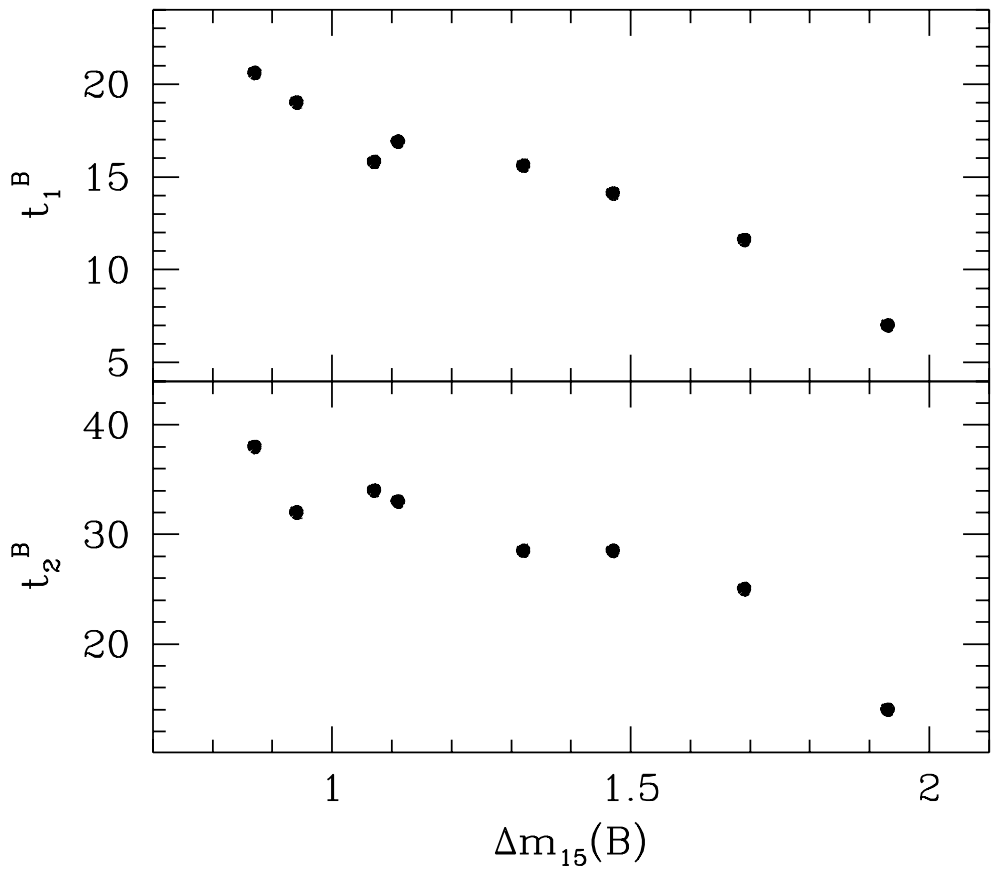}}
\caption{(top) The {\it inflection} point of the B light curve, t$_{1}$$^{B}$,
as a function of $\Delta$m$_{15}$(B). (bottom) The {\it intersection} point of the
B light curve, t$_{2}$$^{B}$, as a function of $\Delta$m$_{15}$(B).}
\end{figure}

\begin{figure}
\psfull
%\psdraft
\centerline{\psfig{figure=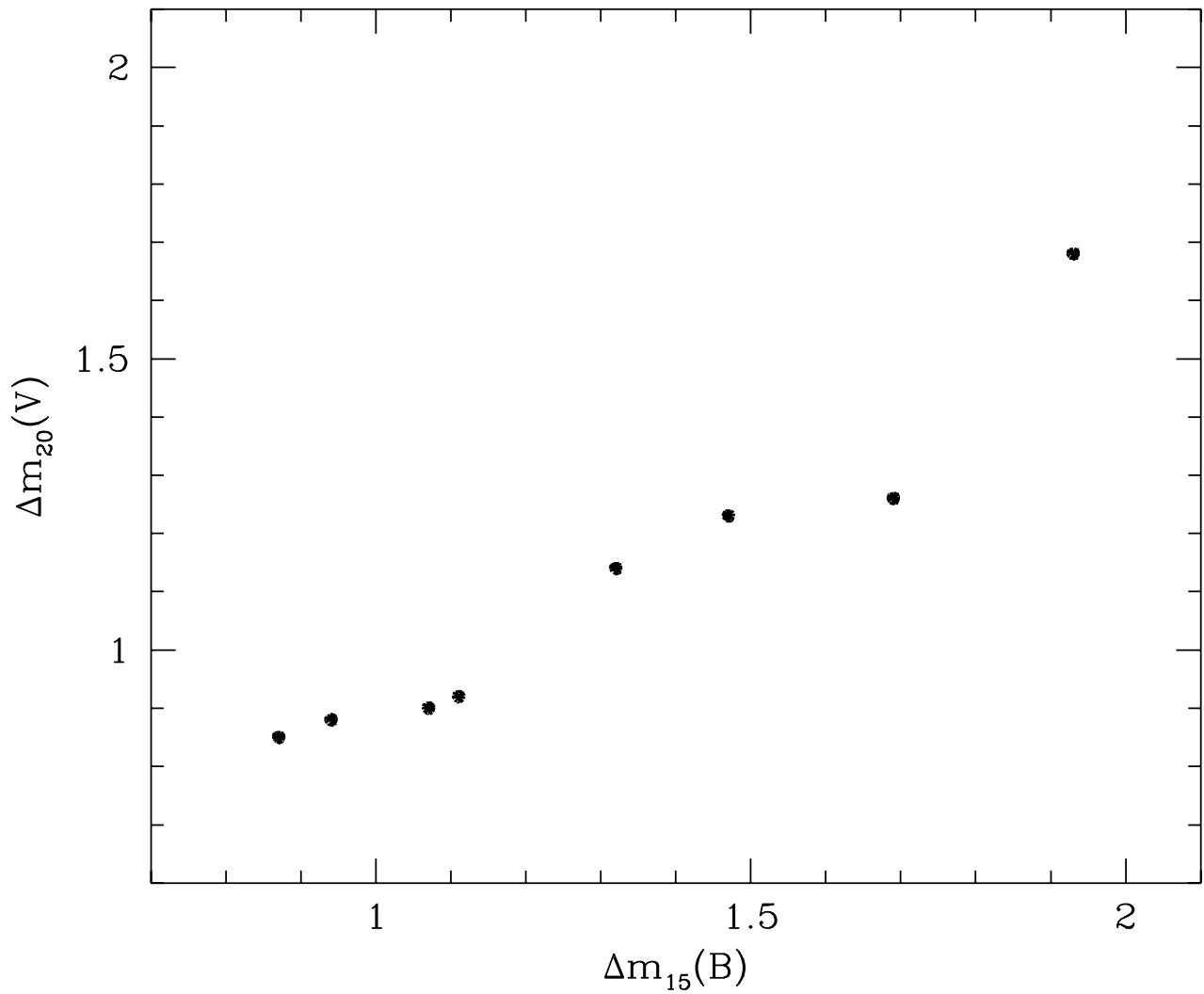}}
\caption{The initial decline rate of the V template light curves, $\Delta$m$_{20}$(V),
plotted as a function of the initial decline rate of the B template light curves, $\Delta$m$_{15}$(B).}
\end{figure}

\begin{figure}
\psfull
%\psdraft
\centerline{\psfig{figure=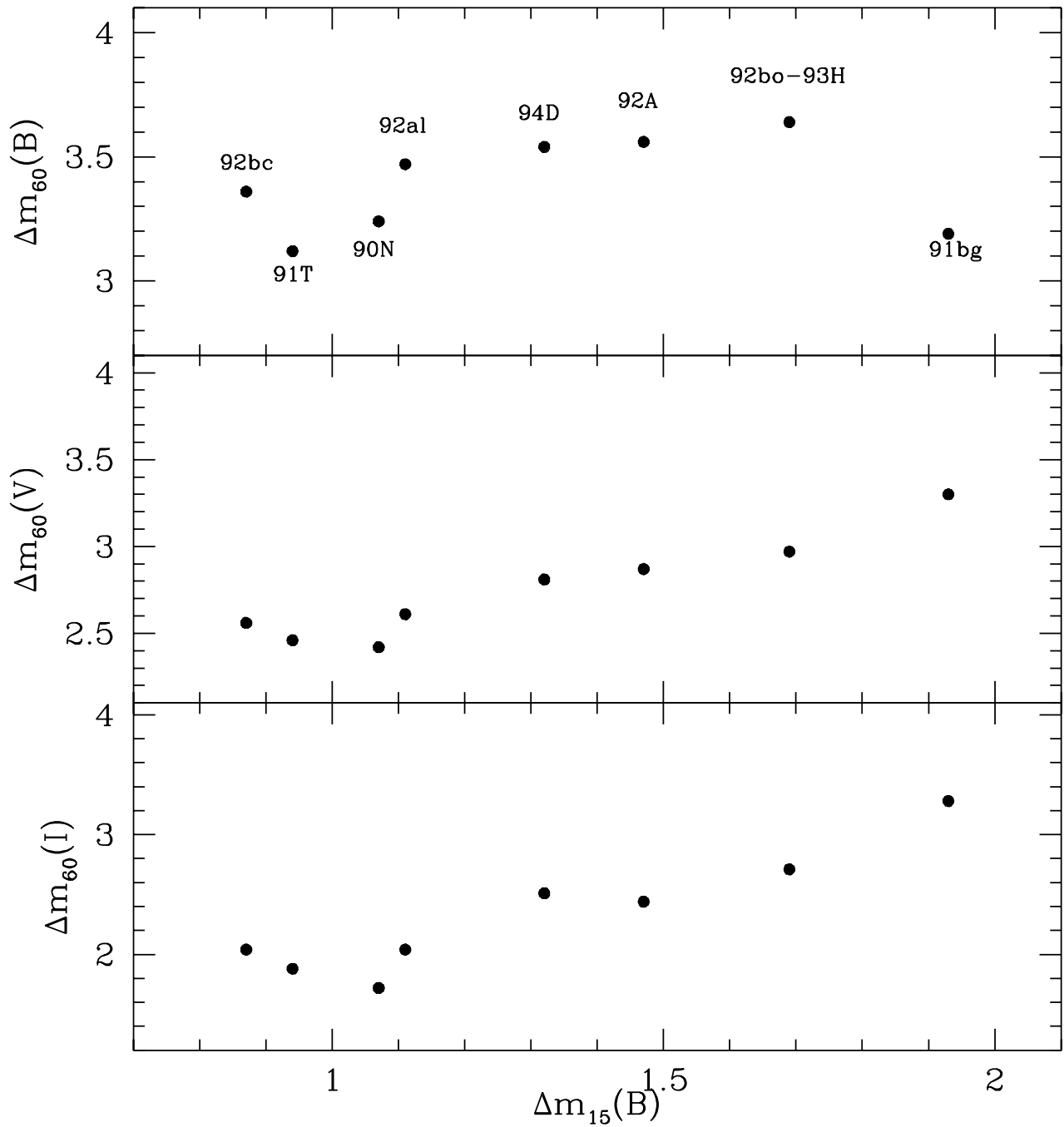}}
\caption{Comparison of the initial decline rate of the B light curve, $\Delta$m$_{15}$(B),
and the brightness of the linear tail (relative to maximum light) of the B (top), V (middle),
and I (bottom) template light curves.}
\end{figure}

\begin{figure}
\psfull
%\psdraft
\centerline{\psfig{figure=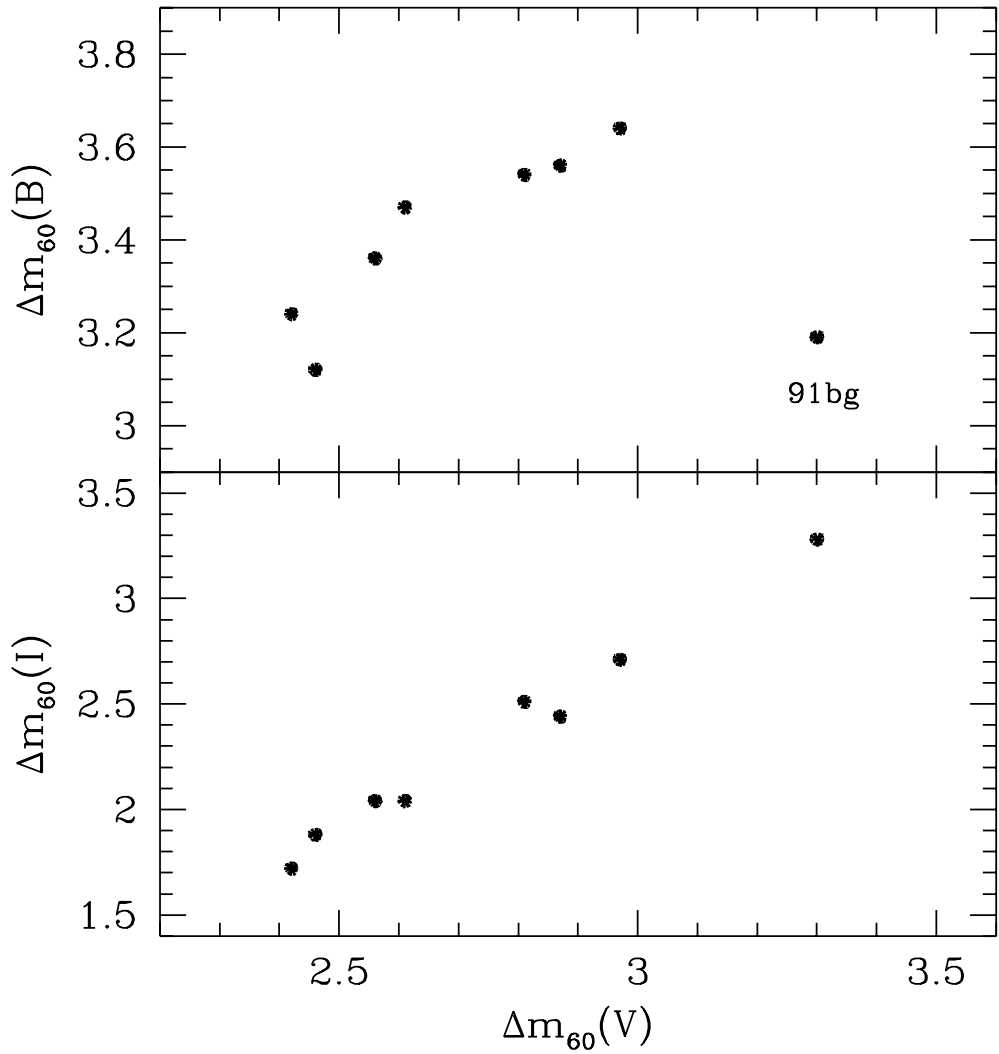}}
\caption{(top) The brightness of the linear tail (relative to maximum light) of the
B template light curves, plotted as a function of the brightness of the linear tail
(relative to maximum light) of the V template light curves. (bottom) The same plot for the
I template curves.}
\end{figure}

\begin{figure}
\psfull
%\psdraft
\centerline{\psfig{figure=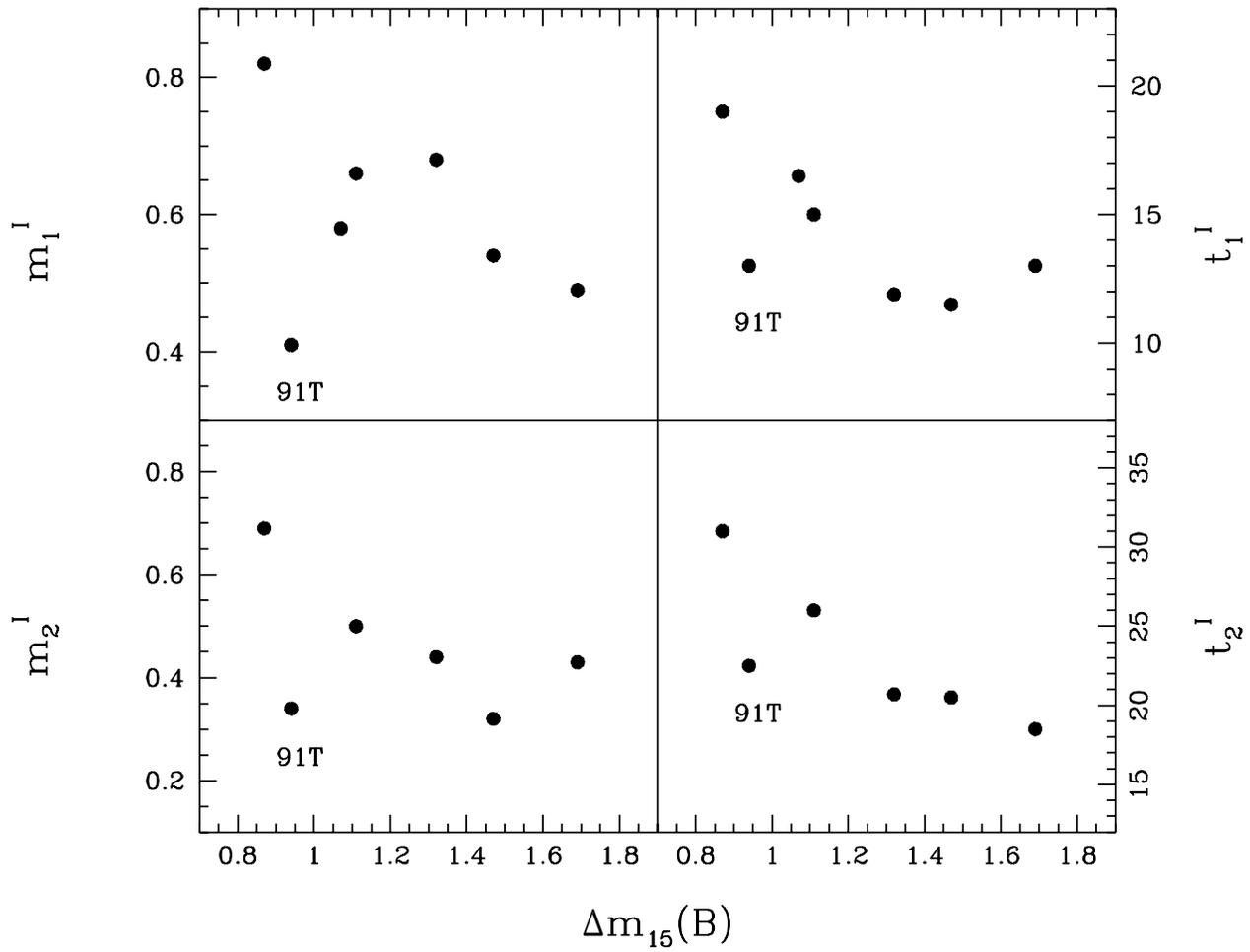}}
\caption{(top) The time and magnitude of the minimum
of the I template light curves, plotted as a function of $\Delta$m$_{15}$(B).
(bottom) The time and magnitude of the secondary maximum
of the I template light curves, plotted as a function of $\Delta$m$_{15}$(B).}
\end{figure}

\begin{figure}
\psfull
%\psdraft
\centerline{\psfig{figure=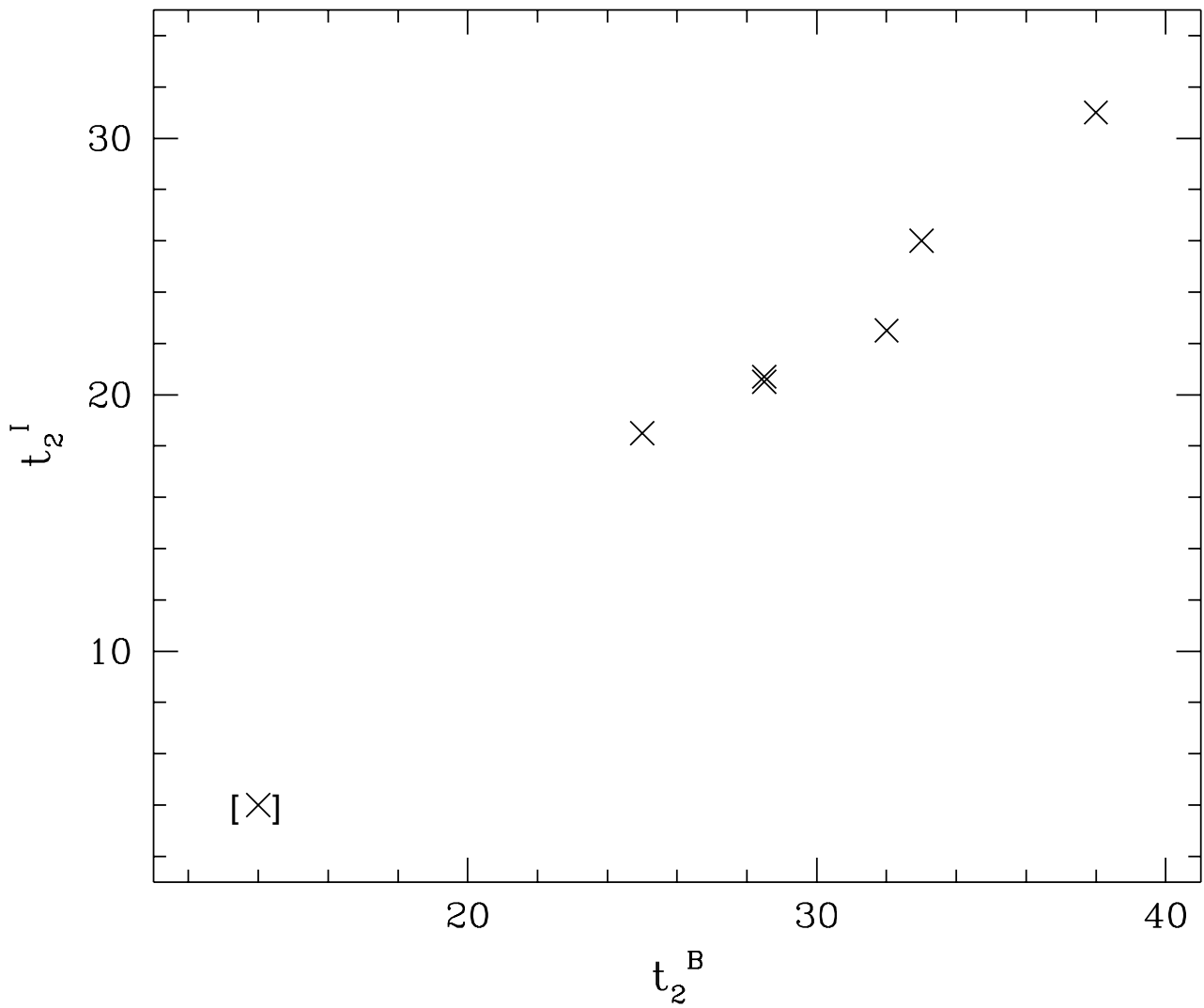}}
\caption{The time of the secondary maximum of the
I template light curves, t$^I_2$, plotted as a function of the time
of occurrence of the bend in the B light curves as measured by the
{\it intersection} parameter, t$^B_2$.}
\end{figure}

\end{document}